\documentclass[preprint,showpacs,preprintnumbers,amsmath,amssymb,aps,floatfix]{revtex4}

\newcommand{\tmpred}{red }
\newcommand{\tmpblue}{blue }
\newcommand{\tmpgreen}{green }
\newcommand{\tmpcyan}{cyan }
\newcommand{\tmpmagenta}{magenta }
\newcommand{\tmpyellow}{yellow }

\usepackage{graphicx}
\usepackage{dcolumn}
\usepackage{bm}

\def\LP{\left(}		
\def\RP{\right)}	

\newcommand{\BE}{\begin{equation}}
\def\EE{\end{equation}}
\def\BEA{\begin{eqnarray}}
\def\EEA{\nonumber\end{eqnarray}}
\def\EL{\nonumber\\}

\def\gtwid{\raise.3ex\hbox{$>$\kern-.75em\lower1ex\hbox{$\sim$}}}
\def\ltwid{\raise.3ex\hbox{$<$\kern-.75em\lower1ex\hbox{$\sim$}}}

\begin{document}

\title{Light hadrons with improved staggered quarks: approaching the continuum limit}

\author{C. Aubin}
\affiliation{Department of Physics, Washington University, St.~Louis, MO 63130, USA}

\author{C. Bernard} 
\affiliation{Department of Physics, Washington University, St.~Louis, MO 63130, USA}



\author{C. DeTar} 
\affiliation{Physics Department, University of Utah, Salt Lake City, UT 84112, USA}

\author{Steven Gottlieb} 
\affiliation{Department of Physics, Indiana University, Bloomington, IN 47405, USA}

\author{E.B. Gregory} 
\affiliation{Department of Physics, University of Arizona, Tucson, AZ 85721, USA}

\author{U.M. Heller} 
\affiliation{American Physical Society, One Research Road, Box 9000,Ridge, NY 11961-9000}

\author{J.E. Hetrick}
\affiliation{University of the Pacific, Stockton, CA 95211, USA}

\author{J. Osborn} 
\affiliation{Physics Department, University of Utah, Salt Lake City, UT 84112, USA}

\author{R. Sugar}
\affiliation{Department of Physics, University of California, Santa Barbara, CA 93106, USA}

\author{D. Toussaint} 
\affiliation{Department of Physics, University of Arizona, Tucson, AZ 85721, USA}

\date{\today}

\begin{abstract}
We have extended our program of QCD simulations with an improved
Kogut-Susskind quark action to a smaller lattice spacing, approximately
0.09 fm.  Also, the simulations with $a \approx 0.12$ fm have been
extended to smaller quark masses.  In this paper we describe the new
simulations and computations of the static quark potential and light
hadron spectrum.  These results give information about the remaining
dependences on the lattice spacing.  We examine the dependence of computed
quantities on the spatial size of the lattice, on the numerical precision
in the computations, and on the step size used in the numerical integrations.
We examine the effects of autocorrelations in ``simulation time'' on the
potential and spectrum.  We see effects of decays, or coupling to two-meson
states, in the $0^{++}$, $1^+$, and $0^-$ meson propagators, and we make
a preliminary mass computation for a radially excited $0^{-}$ meson.
\end{abstract}

\maketitle

%
%
%
%
%
%

\section{Introduction}

We have extended our ongoing program of lattice QCD simulations with three
flavors of dynamical quarks.  In this paper we describe the new simulations we have
done, and present spectrum results for the light hadrons and the static quark
potential.  In a previous work\cite{MILC_spectrum1} we presented results for these
quantities from a set of runs with a lattice spacing of approximately 0.12 fm
and light quark masses ranging down to 0.2 times the estimated strange quark
mass.  Since that time we have extended the $a \approx 0.12$ fm runs to smaller
quark masses, and increased the statistics on the $ m_{u,d} = 0.2 \, m_s$ run.
More importantly, we have done simulations at a smaller lattice spacing of
approximately 0.09 fm in quenched QCD and with three dynamical flavors
at three values of the light quark mass: $ m_{u,d}=m_s$,
$ m_{u,d}=0.4 \, m_s$ and $ m_{u,d}=0.2 \, m_s$ where $m_s$ is the strange quark mass
estimated before doing the simulations\cite{MILC_prelim}.   This enables us to address the
question of lattice spacing effects, {\it i.e.}, extrapolation to the continuum, to
greater accuracy than we could before.
Two short runs were made at larger integration step size than used in the main
simulation as an additional
check on the systematic errors in the simulation algorithm.
At our smallest quark mass, we have computed the hadron propagators in
double precision on a subset of the lattices as a check on the numerical
accuracy of the computations.
Finally, we have done an explicit
test of the effects of the finite spatial size of the simulated system by
adding a run with a larger spatial size than in the main run.

In addition to the light hadron spectrum,
the gluon configurations generated in this program are being used for
computations of
the static quark potential\cite{MILC_potential},
heavy quark and heavy-light meson spectroscopy\cite{HEAVYQ_SPECTRUM,FERMILAB_CHARM},
heavy-light meson decay constants\cite{HEAVYLIGHT_DECAY,FERMILAB_CHARM},
$f_\pi$, $f_K$, and chiral ${\cal O}(p^4)$ parameters \cite{MILC_prelim,MILC_fpi,MILC_fpi_in-prep},
$\alpha_s$\cite{HPQCD_alpha},
exotic meson masses\cite{MILC_exotics},
the topological susceptibility in QCD\cite{MILC_topology},
semileptonic form factors\cite{SEMILEPTONIC_FORM}, 
quark masses\cite{QUARK_MASSES,MILC_fpi,MILC-HPQCD_MASSES}, 
and parton distributions\cite{PARTON_DIST}. 
For those quantities where accurate lattice results are available and
systematic errors are relatively well understood, there is good agreement
with experimental values among a large set of quantities\cite{PRL}.
While this work focuses on describing the simulations, the static potential,
and the light hadron spectrum, results from these other quantities are
important in our analysis.  In particular, the $\Upsilon$ mass splittings
give the most accurate estimates of the lattice spacing, and several
of these quantities enter into our estimates of the correct strange
quark mass.  
In turn, some of the results presented here, such as the dependence of the
static potential on the lattice spacing, and the tests of the effects of 
molecular dynamics step
size and spatial size of the lattices, are important in evaluating these other
works.

\section{Simulations}

\begin{table}[ptbh!]
\begin{center}
\setlength{\tabcolsep}{1.5mm}
\begin{tabular}{|l|l|l|l|l|l|l|l|}
\hline
$am_{u,d}$ / $am_s$  & \hspace{-1.0mm}$10/g^2$  & L & $u_0$ & res. & $\epsilon$ & lats. & $a/r_1$ \\
\hline                             
quenched       & 8.00  & 20 & 0.8879 & na			& na    & 408   & 0.3762(8) \\
\hline                             
0.02  / na     & 7.20  & 20 & 0.8755 & $1\times 10^{-4}$	& 0.013 & 370 & 0.3745(14) \\
\hline                             
0.40  / 0.40   & 7.35  & 20 & 0.8822 & $2\times 10^{-5}$	& 0.03  & 332   & 0.3766(10) \\
0.20  / 0.20   & 7.15  & 20 & 0.8787 & $5\times 10^{-5}$	& 0.03  & 341   & 0.3707(10) \\
0.10  / 0.10   & 6.96  & 20 & 0.8739 & $5\times 10^{-5}$	& 0.03  & 339   & 0.3730(14) \\
0.05  / 0.05   & 6.85  & 20 & 0.8707 & $1\times 10^{-4}$	& 0.02  & 425   & 0.3742(15) \\
0.04  / 0.05   & 6.83  & 20 & 0.8702 & $5\times 10^{-5}$	& 0.02  & 351	& 0.3765(14) \\
0.03  / 0.05   & 6.81  & 20 & 0.8696 & $5\times 10^{-5}$	& 0.02  & 564	& 0.3775(12) \\
0.02  / 0.05   & 6.79  & 20 & 0.8688 & $1\times 10^{-4}$	& 0.0133 & 484	& 0.3775(12) \\
$*$ 0.01  / 0.05   & 6.76  & 20 & 0.8677 & $1\times 10^{-4}$	& 0.00667& 658	& 0.3852(14) \\   
$*$ 0.01  / 0.05   & 6.76  & 28 & 0.8677 & $1\times 10^{-4}$	& 0.00667& 241	& 0.3814(14) \\   
$*$ 0.007  / 0.05   & 6.76  & 20 & 0.8678 & $1\times 10^{-4}$	& 0.005	& 493	& 0.3783(13) \\   
$*$ 0.005  / 0.05   & 6.76  & 24 & 0.8678 & $5\times 10^{-5}$	& 0.003	& 197	& 0.3796(19) \\   
\hline
$*$ quenched       & 8.40  & 28 & 0.8974 & na			& na    & 396   & 0.2681(5) \\
$*$ 0.031  / 0.031   & 7.18  & 28 & 0.8808 & $2\times 10^{-5}$	& 0.02	& 496	& 0.2613(9) \\   
$*$ 0.0124  / 0.031   & 7.11  & 28 & 0.8788 & $5\times 10^{-5}$	& 0.008	& 527	& 0.2698(9) \\   
$*$ 0.0062  / 0.031   & 7.09  & 28 & 0.8782 & $5\times 10^{-5}$	& 0.004	& 592	& 0.2714(9) \\   
\hline
\end{tabular}
\end{center}
\caption{Parameters of the improved action simulations.  A ``*'' at the beginning of
the line indicates a run which is new or has been extended since the report
in Ref~\protect\cite{MILC_spectrum1}.
The first column gives the light and strange quark masses in lattice units, and the second column, the gauge
coupling.  ``L'' is the spatial size of the lattice.  The time size is 64 for the coarse
lattices and 96 for the fine lattices. $u_0$ is obtained from the average
plaquette.   The conjugate gradient residual tabulated here is the residual used in
generating configurations; a smaller residual was used in computing hadron propagators.
``$\epsilon$'' is the time step size in configuration generation.   The second to the
last column is the number of stored lattices, and the last column is the lattice
spacing in units of $r_1$ determined from the static potential in this run.
A ``smoothed'' lattice spacing, discussed later, will be used to convert results to
physical units.
The last four lines, with $a \approx 0.09$ fm, will be referred to as ``fine'' lattices.
\label{RUN_TABLE}
}
\end{table}

The simulations used here are a continuation those described in
Ref.~\cite{MILC_spectrum1}, which contains a more detailed description
of the simulation program.   We use an improved Kogut-Susskind quark
action, the ``$a^2_{\rm tad}$'' or ``Asqtad'' action, which removes lattice artifacts up
to order $a^2 g^2$.  Configurations were generated using the
hybrid-molecular
dynamics ``R~algorithm''\cite{R_ALGORITHM}, with separate pseudofermion
fields for the light and strange quarks, except where all three quarks
are degenerate.   The momenta conjugate to the
gauge fields were refreshed at the end of every trajectory, with the
trajectory length being one simulation time unit.   Lattices were
archived every six time units, and the hadron spectrum and static
quark potential were calculated on these stored lattices. 

Table~\ref{RUN_TABLE} summarizes the parameters of the runs.  For completeness,
it includes runs reported in Ref.~\cite{MILC_spectrum1}, although we will
not repeat tabulation of masses from runs that have not been extended since
that time.  In identifying runs, we will quote the light (degenerate $u$ and $d$)
and strange quark masses as $am_{l/s}=0.01/0.05$, for example.

\section{Static Potential and Length Scale}

\begin{figure}[ptbh!]
\resizebox{6.0in}{!}{\includegraphics{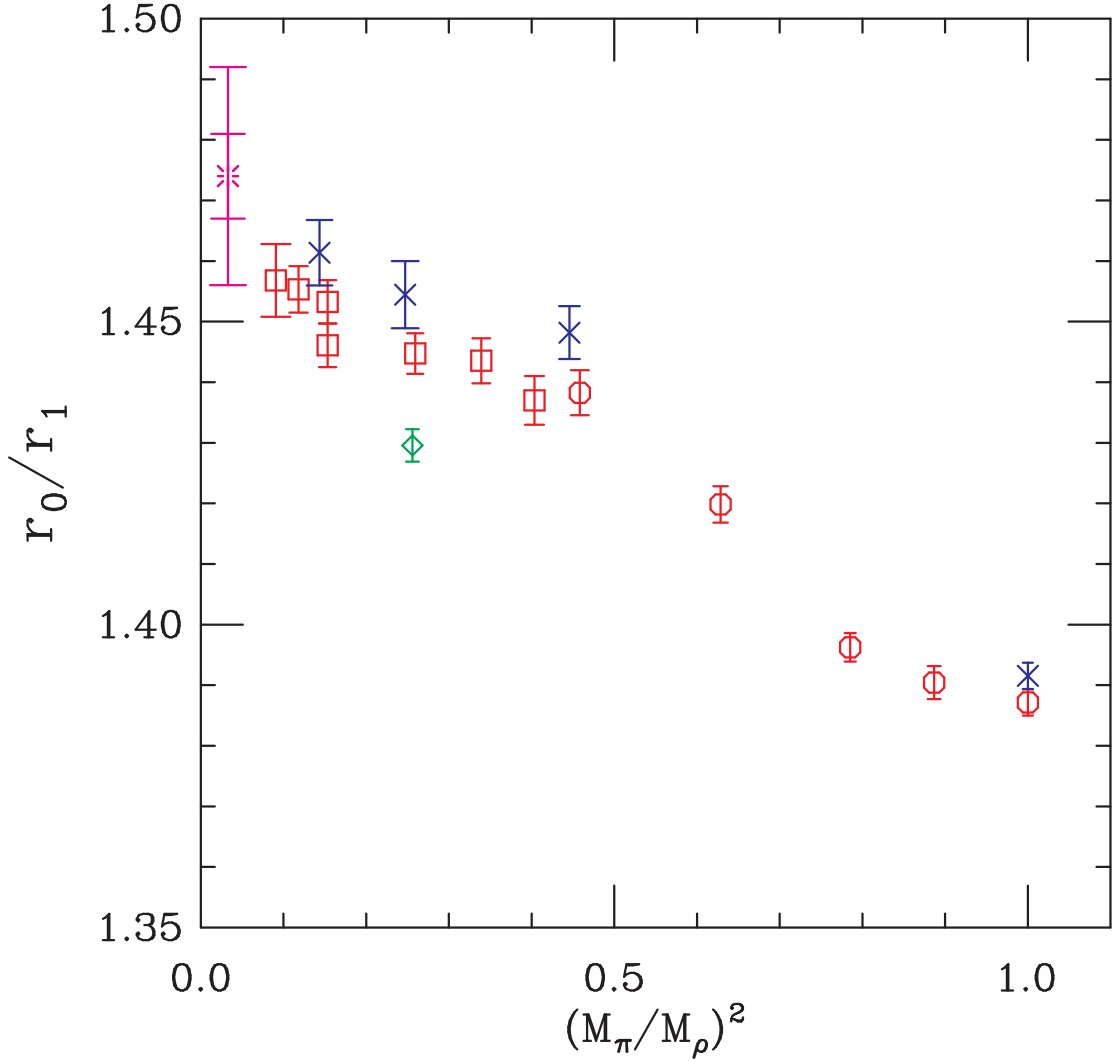}}
\caption{\label{R0_OVER_R1_PLOT}
A ``shape parameter'' for the static potential, $r_0/r_1$.
The \tmpred octagons are from coarse ($a \approx 0.12$ fm) lattices
with three degenerate quark flavors, and the \tmpred squares from
coarse lattices with two light and one strange quark. 
At $(M_\pi/M_\rho)^2=0.15$ the upper square is from the $L=28$ run and
the lower from the $L=20$ run.
The \tmpblue crosses are from the fine ($a \approx 0.09$ fm) runs.
The single \tmpgreen diamond is from a two flavor simulation.
The \tmpmagenta burst is the continuum and chiral extrapolation
discussed in the text, with the smaller error bar the statistical
error and the larger the systematic error.
In this figure we have chosen to use $(M_\pi/M_\rho)^2$ for the
abscissa instead of the $(M_\pi r_1)^2$ used in other figures because
this lets us put the entire range of quark masses up to the quenched
limit ($M_\pi \rightarrow \infty$) in the graph.
}
\end{figure}

\begin{table}[ptbh!]\begin{center}
\begin{tabular}{|l|l|l|l|}
\hline
$am_{u,d}$ / $am_s$  & \hspace{-1.0mm}$10/g^2$ & $r_1/a$ (run) & $r_1/a$ (smoothed) \\
\hline
0.0492/0.082	& 6.503	& 1.774(10)	& 1.778 \\
0.0328/0.082	& 6.485	& 1.786(10)	& 1.788 \\
0.0164/0.082	& 6.467	& 1.783(12)	& 1.797 \\
0.0082/0.082	& 6.458	& 1.807(10)	& 1.802 \\
0.082/0.082	& 6.561	& 1.816(10)	& 1.805 \\
0.0492/0.0492  & 6.475   & 1.807(28)	& 1.766 \\
0.0328/0.0328  & 6.470   & 1.768(30)	& 1.828 \\
0.0164/0.0164  & 6.430   & 1.796(22)	& 1.813 \\
0.0492/0.0492  & 6.500   & 1.818(23)	& 1.821 \\
0.0492/0.0492  & 6.450   & 1.735(30)	& 1.713 \\
0.0328/0.0328  & 6.450   & 1.757(30)	& 1.784 \\
0.0164/0.0164  & 6.450   & 1.857(25)	& 1.858 \\
0.0082/0.0082  & 6.420   & 1.843(20)	& 1.827 \\
\hline
0.005/0.050	& 6.76	& 2.634(13)	& 2.632 \\
0.007/0.050	& 6.76	& 2.644(09)	& 2.623 \\
0.010/0.050	& 6.76	& 2.598(08)	& 2.610 \\
0.010/0.050	& 6.76	& 2.621(09)	& 2.610 \\
0.020/0.050	& 6.79	& 2.649(08)	& 2.650 \\
0.030/0.050	& 6.81	& 2.656(10)	& 2.662 \\
0.040/0.050	& 6.83	& 2.666(11)	& 2.673 \\
0.050/0.050	& 6.85	& 2.679(11)	& 2.683 \\
0.030/0.030	& 6.79	& 2.678(14)	& 2.650 \\
\hline
0.031/0.031	& 7.18	& 3.827(12)	& 3.822 \\
0.0124/0.031	& 7.11	& 3.707(13)	& 3.711 \\
0.0062/0.031	& 7.09	& 3.687(12)	& 3.684 \\
\hline
\end{tabular}
\end{center}
\caption{
Smoothed $r_1/a$ compared with $r_1/a$ determined from
each run.  The top block is from lattices with $a \approx 0.18$ fm
from tuning runs for our high temperature simulations, while the
second and third blocks are the ``coarse'' and ``fine'' lattices respectively.
Five short ``tuning runs'' are omitted from this table.
Several of the runs have been extended since fitting of the smoothed
$r_1$ was done.
\label{SMOOTH_TABLE}
} \end{table}

We use the static quark potential to relate the lattice spacings in our different
runs.  In particular, we use the quantity  $r_1$ defined by $r_1^2 F(r_1)=1.00$.
We choose $r_1$ because of its ease and accuracy of computation and lack of dependence
on the valence quark mass.  Computation of this quantity and the effects of dynamical
quarks on the potential have been discussed in Refs.~\cite{MILC_potential,MILC_spectrum1}.
Here we add points at smaller quark mass and, more importantly, points at a finer lattice
spacing which allow a preliminary continuum extrapolation.
As before, we fit to the form in Ref.~\cite{UKQCD_POTFORM},
\BE V(\vec r) = C + \sigma r - \alpha/r + \lambda \LP V_{\rm free}(\vec r) - 1/r \RP \ \ . \EE
where  $V_{\rm free}(\vec r)$ is the potential calculated in
free field theory, using the improved gauge action.  This lattice correction term
is used at distances less than $3a$.
%

While we expect $r_1/a$ to be a smooth function of the quark masses and gauge
couplings, $r_1/a$ determined from fitting the potential in a particular run
will have a statistical error, and fluctuate from its ideal (infinite statistics)
value.   To minimize the effects of these run-to-run fluctuations, we have
fit a smoothed $r_1/a$ for our three flavor lattices with quark masses
less than or equal to the strange quark mass.  Over the range of masses
and gauge couplings we have used, a simple fitting form
\BE \log(r_1/a) = C_{00} + C_{10} \LP \frac{10}{g^2}-7.0 \RP
   + C_{01} \LP 2m_{u,d}+m_s \RP + C_{20} \LP \frac{10}{g^2}-7.0 \RP^2 \EE
gives an acceptable fit with a $\chi^2$ of 30.3 with 26 degrees of
freedom, with
\BEA
  C_{00} &=& 1.2578(27) \EL
  C_{10} &=& 0.9371(93) \EL
  C_{01} &=& -0.828(29) \EL
  C_{20} &=& -0.271(22) \\
\EEA
Table~\ref{SMOOTH_TABLE} shows values of $r_1/a$ used in the fit
together with the smoothed $r_1/a$ for each run.
We have used this smoothed $r_1/a$ in converting results from units
of the lattice spacing into units of $r_1$.

The shape of the static quark potential is affected by dynamical quarks.
One of many possible ratios parameterizing this shape is the ratio $r_0/r_1$.
We use the results in Fig.~\ref{R0_OVER_R1_PLOT} to extrapolate $r_0/r_1$
to the physical quark mass and continuum limit.  Simultaneously fitting
coarse and fine lattice results to a constant plus linear terms in the
quark mass and $a^2\,\alpha_s$ gives
\BE r_0/r_1 = 1.476(7) - 0.049(10)(M_\pi/M_\rho)^2 - 0.12(4)(a/r_1)^2\,\alpha_s(a)/\alpha_s(0.12 {\rm fm}) \ \ \ ,\EE
with $\chi^2=3.6$ for 8 degrees of freedom, using $\alpha_s$ from Ref.~\cite{HPQCD_alpha}.
In fitting the potential the same distance range, $\sqrt{2}-6$, was used for
all the coarse lattices, and range $\sqrt{5}-7$ for all the fine lattices.
Therefore, the statistical error bars in Table~\ref{SMOOTH_TABLE} and
Fig.~\ref{R0_OVER_R1_PLOT} appropriately represent the fluctuations in $r_1/a$
or $r_0/r_1$ within each of these two sets of runs.  However, there is
a systematic effect from the choice of fit range which is common to all
coarse runs and all fine runs, but may differ between the two sets.
Varying the fitting range over reasonable ranges suggests that this
systematic error can be conservatively estimated as an uncertainty of
0.01 in the difference between the coarse and fine lattice $r_0/r_1$.
This leads to a systematic uncertainty of about 0.018 in the continuum
extrapolation, leading to an estimate
\BE r_0/r_1 =  1.474(7)(18) \EE
at the physical $M_\pi/M_\rho$ in the continuum limit.

To compute $r_1$ in physical units, we need to set the lattice
scale using a directly measurable physical quantity.  A convenient choice is
the $\Upsilon$ spectrum, in particular the 2S--1S and 1P--1S splittings.
This gives a scale $a^{-1}=1.588(19)$ GeV on the coarse $0.01/0.05$ lattices,
and $a^{-1}=2.271(28)$ GeV on the fine $0.0062/0.031$ lattices \cite{HPQCD_private}.
For light quark masses $\ltwid m_s/2$, the mass dependence of these quantities
and of $r_1$ appears to be slight, and we neglect it.  With our smoothed values
of $r_1/a$, we then get $r_1= 0.324(4)$ fm on the coarse lattices and
$r_1= 0.320(4)$ fm on the fine lattices.

To extrapolate $r_1$ to the continuum, we first
assume that the dominant discretization errors go like $\alpha_S a^2$.  Using
$\alpha_V(q^*)$ \cite{HPQCD_alpha,LEPAGE_MACKENZIE} (with scale $q^*=3.33/a$) for $\alpha_S$
gives a ratio $(\alpha_S a^2)_{\rm fine}/(\alpha_S a^2)_{\rm coarse}=0.428$.
Extrapolating away the discretization errors linearly then results in
$r_1=0.317(7)$ fm in the continuum.  However, taste-violating effects,
while formally ${\cal O}(\alpha^2_S a^2)$ and hence subleading, are known to
be at least as important as the leading errors in some cases.  Therefore, one should
check if the result changes when the errors are assumed to go like $\alpha^2_S a^2$.
Taking $\alpha_S=\alpha_V(3.33/a)$ gives a ratio
$(\alpha^2_S a^2)_{\rm fine}/(\alpha^2_S a^2)_{\rm coarse}=0.375$; while
a direct lattice measurement of the taste-splittings
to be presented in the next section
gives a ratio of $0.35$.
Extrapolating linearly to the continuum then implies
$r_1=0.318(7)$ fm or $r_1=0.319(6)$ fm respectively, in agreement with the
previous result.  For our final result, we use an ``average'' ratio of 0.4 and
add the effect of varying this ratio in quadrature with the statistical error.
We obtain $r_1=0.317(7)(3)$ fm.  The second error is a crude estimate of the
systematic error from the choice of fit ranges for the static potential.

A similar calculation to estimate $r_0$ yields $0.471(6)$ fm on the coarse run and
$0.466(6)$ fm on the fine run, with a continuum extrapolated value of $0.462(11)(4)$ fm,
where the second error is an estimate of the systematic error from choice of
fit ranges in the potential.  If we take the above estimate of $r_0/r_1$ and
multiply by $r_1=0.317$ fm, we obtain instead $r_0=0.467$ fm, and the difference
in these two calculations of $r_0$ is another measure of systematic error.

\section{Light Hadron Masses}

Our procedures for calculating and fitting hadron propagators are described in
Ref.~\cite{MILC_spectrum1}.  With the exception of the non-Goldstone pions
at $am_{u,d}=0.0124$, we used Coulomb gauge wall sources, with eight source
time slices evenly spread through the lattice.
Propagators were fit with varying minimum distances, and with the maximum
distance either at the midpoint of the lattice or where the fractional statistical
errors exceeded 30\% for two successive time slices.
In most cases, to reduce the effect of autocorrelations,
propagators from four successive lattices (24 simulation time units)
were blocked together before computing the covariance matrix.
Masses were selected by looking for a combination of a ``plateau'' in the mass
as a function of minimum distance and a good confidence level ($\chi^2$) for the
fit.  We also made an effort to choose minimum distances that are smooth functions
of the couplings, recognizing that statistically we should have some fits with low
and high confidence levels.   

\subsection{Pseudoscalar mesons}\label{ps_mesons}

\begin{figure}[ptbh!]
\resizebox{6.0in}{!}{\includegraphics{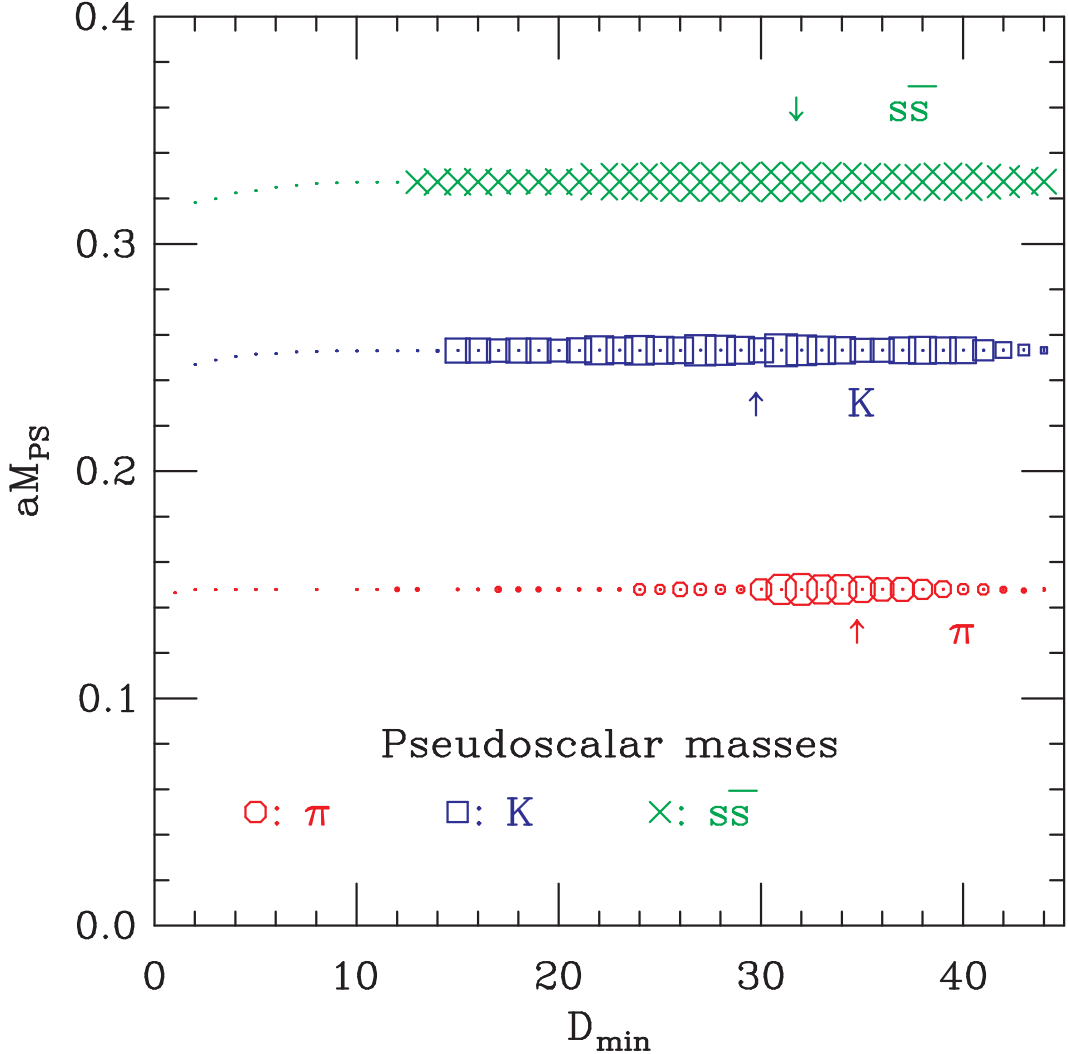}}
\caption{\label{dmin_ps_0062_fig}
Pseudoscalar masses as a function of minimum distance included in the fit
from the run with $10/g^2=7.09$ and $a m_{l/s}=0.0062/0.031$.
The size of the symbols is proportional to the confidence level of the fit,
with the size of the symbols in the labels corresponding to 50\%.
These fits included only a single exponential.
Fits selected to quote in the mass tables are marked with arrows.
}
\end{figure}

We calculated masses for the exact Goldstone ($\gamma_5 \otimes \gamma_5$) pseudoscalar
mesons in all of the runs.   For the $a m_{l/s}=0.0124/0.031$ run we calculated the masses
of all of the different taste pions, allowing us to see how the taste symmetry breaking
decreases with lattice size.  Figure~\ref{dmin_ps_0062_fig} shows the fitted masses for
the pion, the kaon and the ``unmixed $s \bar s$'' from the fine lattice run with
$m_{u,d}=0.2 \, m_s$.
Table~\ref{PS_MASS_TABLE} shows the selected fits for the pseudoscalar meson masses.

\begin{table}[ptbh!]
\begin{center}
\begin{tabular}{|llllll|}
\hline
$am_{valence}$ & $am_{sea}$ & $am_{PS}$ & range & $\chi^2/D$ & conf. \\
\hline
0.015 ($\pi$)   & $\infty$      & 0.21643(14)   & 18--47 & 25/28 & 0.62 \\
0.03 ($\pi$)    & $\infty$      & 0.30259(14)   & 24--47 & 21/22 & 0.53 \\
\hline
0.01 ($\pi$)    & 0.01/0.05     & 0.22439(20)   & 19--31 & 9.1/11        & 0.61 \\
0.01 ($\pi$)    & 0.01/0.05     & 0.22421(12)   & 19--31 & 4.7/11        & 0.94 \\
0.007 ($\pi$)   & 0.007/0.05    & 0.18881(19)   & 20--31 & 14/10 & 0.18 \\
0.005 ($\pi$)   & 0.005/0.05    & 0.15970(20)   & 22--31 & 11/8  & 0.19 \\
0.01/0.05 ($K$) & 0.01/0.05     & 0.38327(22)   & 17--32 & 23/14 & 0.067 \\
0.01/0.05 ($K$) & 0.01/0.05     & 0.38304(20)   & 17--32 & 14/13 & 0.38 \\
0.007/0.05 ($K$)        & 0.007/0.05    & 0.37268(25)   & 20--31 & 8.6/10        & 0.57 \\
0.005/0.05 ($K$)        & 0.005/0.05    & 0.36550(29)   & 20--31 & 6.4/10        & 0.78 \\
0.05 ($s\bar s$)        & 0.01/0.05     & 0.49427(18)   & 17--32 & 19/14 & 0.18 \\
0.05 ($s\bar s$)        & 0.01/0.05     & 0.49443(18)   & 17--31 & 17/13 & 0.20 \\
0.05 ($s\bar s$)        & 0.007/0.05    & 0.49317(19)   & 20--31 & 12/10 & 0.31 \\
0.05 ($s\bar s$)        & 0.005/0.05    & 0.49276(23)   & 20--31 & 5.2/10        & 0.87 \\
\hline
0.031 ($\pi$)   & 0.031/0.031   & 0.32003(18)   & 25--47 & 20/21 & 0.52 \\
0.0124 ($\pi$)  & 0.0124/0.031  & 0.20638(18)   & 30--47 & 22/16 & 0.15 \\
0.0062 ($\pi$)  & 0.0062/0.031  & 0.14794(19)   & 35--47 & 7/11  & 0.8 \\
0.0124/0.031 ($K$)      & 0.0124/0.031  & 0.27209(18)   & 30--47 & 23/16 & 0.11 \\
0.0062/0.031 ($K$)      & 0.0062/0.031  & 0.25319(19)   & 30--47 & 14/16 & 0.61 \\
0.031 ($s\bar s$)       & 0.0124/0.031  & 0.32585(17)   & 27--47 & 29/19 & 0.07 \\
0.031 ($s\bar s$)       & 0.0062/0.031  & 0.32727(14)   & 32--47 & 5.6/14        & 0.97 \\

\hline
\end{tabular}
\caption{
\advance\baselineskip -6pt
Pseudoscalar meson masses.
Here we include runs that are new or have been extended since Ref.~\protect\cite{MILC_spectrum1};
results at larger quark masses can be found there.
The first column is the valence quark mass(es), and the
second column the sea quark mass or masses.
The particle name is in the first column.  Here
``$\pi$'' indicates valence quark mass equal to the lighter dynamical
quarks, or degenerate in the quenched case. ``$K$'' indicates one valence
quark equal to the light dynamical quarks and one at about $m_s$, while
``$s \bar s$''  indicates a fictitious meson with
two valence quarks with mass about $m_s$, in a flavor nonsinglet state.
The remaining columns are the hadron mass, the time range for the
chosen fit, $\chi^2$ and number of degrees of freedom for the
fit, and the confidence level of the fit.
The first block is from the quenched run at $10/g^2=8.4$, the second block from
the coarse three flavor runs, and the last block from the fine three flavor runs.
The two lines with $am_{sea}=0.01/0.05$ are from the runs with $L=20$ and $28$.
\label{PS_MASS_TABLE}
}
\end{center}\end{table}

With Kogut-Susskind quarks there are four ``tastes'' of valence quark, and
hence sixteen different tastes of pseudoscalar mesons, grouped in eight
multiplets.  In the continuum limit
these are degenerate, and the improved action reduces these splittings
relative to the one-link fermion action.
In our previous work on the coarse lattices we verified that these
pion masses show the partial taste symmetry restoration predicted
by Lee and Sharpe\cite{PARTIAL_FLAVOR_SYM}.  In particular,
we expect near degeneracy between pairs of pions between which
$\gamma_0$ is replaced by $\gamma_i$, {\it e.g.} taste $\gamma_0\gamma_5$ with
taste $\gamma_i\gamma_5$.
Also, the squared masses are approximately linear in the quark mass, with all tastes having
the same slope.   This means that a dimensionless measure of taste
symmetry breaking, $\LP M_\pi^2-M_G^2 \RP r_1^2$, is almost independent
of the quark mass.
Having verified these properties on the coarse lattice, we computed non-pointlike
pion propagators on only one of the fine lattice runs, with $10/g^2=7.11$ and
$a m_{l/s} = 0.0124/0.031$, which has a lattice spacing of $a/r_1=0.269$.  In 
Table~\ref{PION_SPLIT_TABLE} we give these pion masses, together with those
from the coarse lattice run with comparable quark masses.  To facilitate
comparison, these masses are given in units of $r_1$.  We also give
the measure of taste symmetry breaking, $\LP M_\pi^2-M_G^2 \RP r_1^2$, for
these masses.   It can be seen that $\LP M_\pi^2-M_G^2 \RP r_1^2$ for each
taste on the fine lattices is consistently about 0.35 times the value on
the coarse lattices.   This is consistent with the expected scaling as
$a^2 \alpha_S^2$ described above, which, using $\alpha_s=\alpha_V(q^*)$ and $q^*=3.33/a$\ \cite{HPQCD_alpha}
suggests a ratio of 0.375.

\begin{table}[ptbh!]\begin{center}
\begin{tabular}{|c|c|c|c|c|c|}
\hline
pion taste	& $M_\pi r_1$ (coarse)	& $\LP M_\pi^2-M_G^2 \RP r_1^2$  (coarse) &
 $M_\pi r_1$ (fine)  & $\LP M_\pi^2-M_G^2 \RP r_1^2$  (fine) & ratio \\

\hline
$\gamma_5$		& 0.8251(45)	& - & 0.7659(7)	& - & - \\
$\gamma_0 \gamma_5$	& 0.9386(19)	& 0.2003(35) & 0.8127(11)	& 0.0739(18) & 0.369(11)\\
$\gamma_i \gamma_5$	& 0.9426(16)	& 0.2078(30) & 0.8116(26)	& 0.0721(42) & 0.347(21)\\
$\gamma_i \gamma_j$	& 1.0033(34)	& 0.3259(69) & 0.8372(41)	& 0.1143(68) & 0.351(22)\\
$\gamma_i \gamma_0$	& 1.0044(29)	& 0.3280(59) & 0.8383(26)	& 0.1162(44) & 0.354(15)\\
$\gamma_i$		& 1.0555(53)	& 0.4334(12) & 0.8576(56)	& 0.1489(95) & 0.344(22)\\
$\gamma_0 $		& 1.0558(32)	& 0.4339(67) & 0.8602(37)	& 0.1534(64) & 0.354(16)\\
${\bf 1}$		& 1.1029(80)	& 0.5358(75) & 0.8899(93)	& 0.2054(165) & 0.383(31)\\
\hline
\end{tabular}
\caption{
Taste symmetry violations on coarse and fine lattices.  The second and fourth
columns contain the masses for the different pions in units of $r_1$ for
a coarse and fine lattice run.
The coarse lattice run (from Ref. ~\protect\cite{MILC_spectrum1}) was at
$10/g^2=6.79$ and $a m_{l/s}=0.02/0.05$, and had a lattice spacing $a/r_1=0.377$. 
The fine lattice run was at
$10/g^2=7.11$ and $a m_{l/s}=0.0124/0.031$, and had a lattice spacing $a/r_1=0.269$. 
The physical quark masses are similar, as evidenced by the similar Goldstone
pion masses.
The third and fifth columns are a measure of taste symmetry breaking,
$\LP M_\pi^2-M_G^2 \RP r_1^2$, on the coarse and fine lattices, and the
final column is the ratio of this measure between the fine and coarse lattice runs.
}
\label{PION_SPLIT_TABLE}
\end{center}\end{table}

In a separate analysis we calculate
``partially quenched'' pseudoscalar masses and decay constants,
where the valence quark and sea quarks have different
masses\cite{MILC_fpi,MILC_fpi_in-prep}.  These results have
been analyzed using chiral perturbation theory including terms parameterizing the
taste symmetry breaking\cite{SCHPT}.  
From this analysis we find $f_\pi$ and $f_K$ at the physical quark masses,
and values for several of the low energy constants in chiral perturbation theory.
Another product of the computations of $m_{PS}$ and $f_{PS}$ is a determination
of the lattice quark masses corresponding to the real world.
We define the strange and light quark masses at fixed lattice spacing,
$am^{\rm lat}_s$ and $am^{\rm lat}_{u,d}$, to be the lattice masses that
give the experimental values for $M_K$ and $M_\pi$.  To determine
$am^{\rm lat}_s$ and $am^{\rm lat}_{u,d}$, we fit the mass and decay constant
data to chiral log forms that take into account staggered
taste violations \cite{SCHPT}. We find
$am^{\rm lat}_s = 0.0390(1)({}^{+18}_{-20})$,
$am^{\rm lat}_{u,d}= 0.00148(1)({}^{+6}_{-8})$ on the coarse lattices, and
$am^{\rm lat}_s = 0.0272(1)({}^{+12}_{-10})$,
$am^{\rm lat}_{u,d}= 0.00103(0)(4)$ on the fine lattices,
where the errors are statistical and systematic.
The systematic error is dominated by that coming from
the chiral extrapolation/interpolation and the $\sim\!2\%$
scale uncertainty.

We have also calculated masses of excited pseudoscalar mesons.
Because this requires consideration of two-meson states,
discussion of this is deferred to a later section on hadronic decays 
and excited states.

\subsection{Vector mesons}

\begin{figure}[ptbh!]
\resizebox{6.0in}{!}{\includegraphics{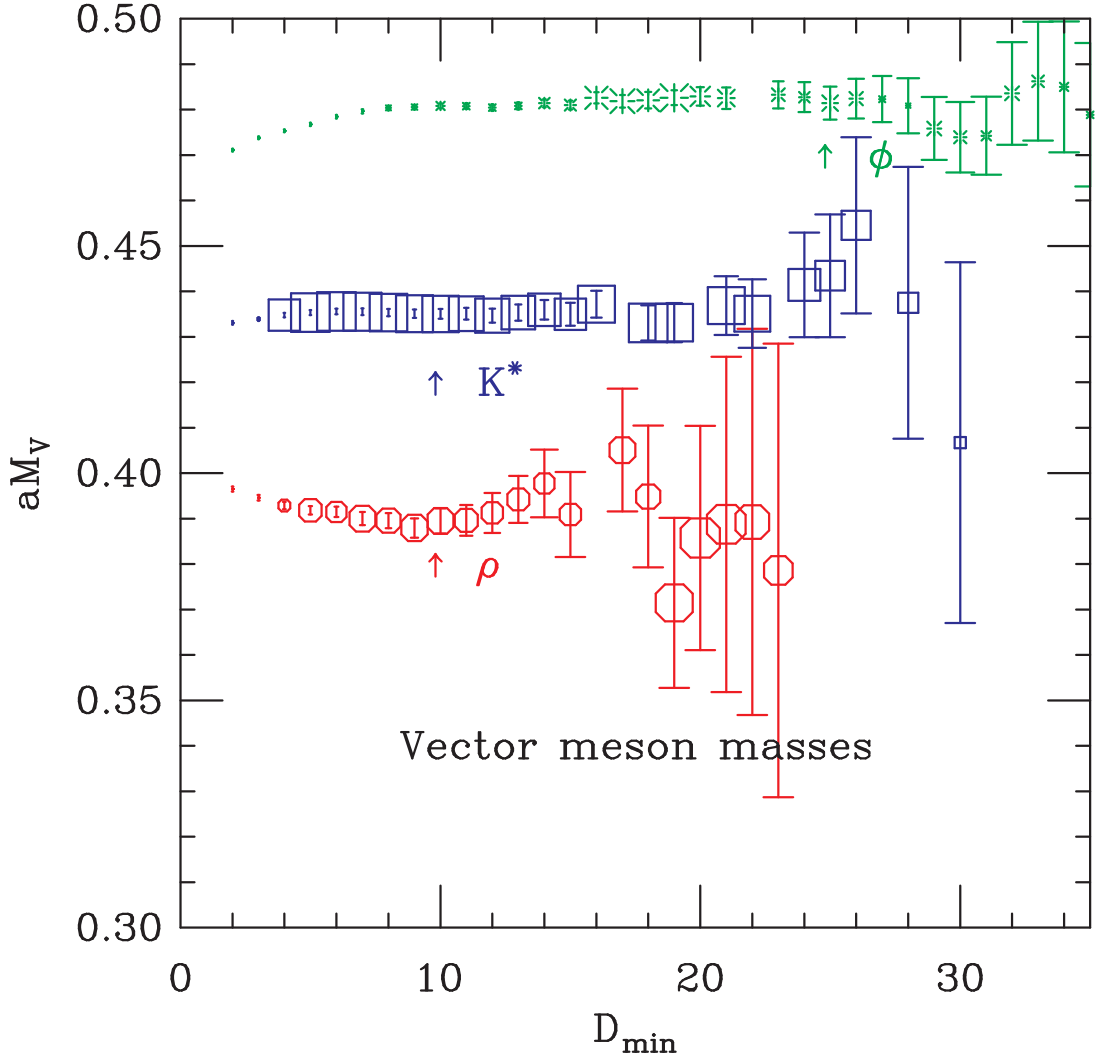}}
\caption{\label{dmin_vt_0062_fig}
Vector meson masses as a function of minimum distance included in the fit
from the run with $10/g^2=7.09$ and $a m_{l/s}=0.0062/0.031$.
}
\end{figure}

Figure~\ref{dmin_vt_0062_fig} shows vector meson masses {\it versus}
minimum distance fit for the fine lattice run with the lightest quark mass.
Mass estimates for all of the runs are in Table~\ref{VECTOR_MASS_TABLE}.
Note that despite our relatively small quark masses, none of these
vector mesons are below the threshold for decay into two pseudoscalars,
since the angular momentum of the vector mesons requires that the vector
meson at rest decay into pseudoscalars with momentum $2\pi/L$.
In addition, we require a combination of tastes in the pseudoscalars
that overlaps with the taste of the vector meson --- the vector mesons
tabulated here have spin $\otimes$ taste $ = \gamma_i \otimes \gamma_i$.

Table~\ref{PSEUDOVECTOR_MASS_TABLE} shows masses for $1^+$ mesons.
These mesons can decay into a vector and a pseudoscalar meson, and these
simulations reach into the quark mass region where this threshold is
crossed.  We defer discussion of this effect to the next section.

\begin{table}[ptbh!]
\begin{center}
\begin{tabular}{|llllll|}
\hline
$am_{valence}$ & $am_{sea}$ & $am_{V}$ & range & $\chi^2/D$ & conf. \\
\hline
0.015 ($\rho$)  & $\infty$      & 0.4660(30)    & 11--25 & 4/11  & 0.97 \\
0.03 ($\rho$)   & $\infty$      & 0.4992(15)    & 5--25  & 18/15 & 0.28 \\
\hline
0.01 ($\rho$)   & 0.01/0.05     & 0.5690(50)    & 6--22  & 15/13 & 0.32 \\
0.01 ($\rho$)   & 0.01/0.05     & 0.5680(30)    & 6--19  & 10/10 & 0.42 \\
0.007 ($\rho$)  & 0.007/0.05    & 0.5510(40)    & 6--18  & 11/9 & 0.26 \\
0.005 ($\rho$)  & 0.005/0.05    & 0.5340(80)    & 6--15  & 5.8/6 & 0.44 \\
0.01/0.05 ($K^*$)       & 0.01/0.05     & 0.6492(25)    & 8--23  & 5.2/12        & 0.95 \\
0.01/0.05 ($K^*$)       & 0.01/0.05     & 0.6462(18)    & 8--27  & 29/16 & 0.023 \\
0.007/0.05 ($K^*$)      & 0.007/0.05    & 0.6330(30)    & 9--23  & 10/11 & 0.54 \\
0.005/0.05 ($K^*$)      & 0.005/0.05    & 0.6190(40)    & 10--23 & 15/10 & 0.12 \\
0.05 ($\phi$)   & 0.01/0.05     & 0.7193(14)    & 9--30  & 11/18 & 0.90 \\
0.05 ($\phi$)   & 0.01/0.05     & 0.7194(11)    & 9--31  & 15/19 & 0.74 \\
0.05 ($\phi$)   & 0.007/0.05    & 0.7114(16)    & 12--30 & 12/15 & 0.69 \\
0.05 ($\phi$)   & 0.005/0.05    & 0.7140(30)    & 14--29 & 15/12 & 0.25 \\
\hline
0.031 ($\rho$)  & 0.031/0.031   & 0.4781(14)    & 16--42 & 36/23 & 0.043 \\
0.0124 ($\rho$) & 0.0124/0.031  & 0.4173(13)    & 10--33 & 31/20 & 0.059 \\
0.0062 ($\rho$) & 0.0062/0.031  & 0.3895(28)    & 10--27 & 11/14 & 0.65 \\
0.0124/0.031 ($K^*$)    & 0.0124/0.031  & 0.4483(18)    & 15--42 & 42/24 & 0.013 \\
0.0062/0.031 ($K^*$)    & 0.0062/0.031  & 0.4350(11)    & 10--34 & 13/21 & 0.91 \\
0.031 ($\phi$)  & 0.0124/0.031  & 0.4831(8)     & 14--47 & 55/30 & 0.0032 \\
0.031 ($\phi$)  & 0.0062/0.031  & 0.4810(40)    & 25--45 & 18/17 & 0.39 \\
\hline
\end{tabular}
\caption{Vector meson masses.
Runs tabulated and the format are the same as in Table~\protect\ref{PS_MASS_TABLE}.
Here ``$\rho$'' indicates valence quark mass equal to the lighter dynamical
quarks, or degenerate in the quenched case. ``$K^*$'' indicates one valence
quark equal to the light dynamical quarks and one at about $m_s$, while
``$\phi$''  indicates two valence quarks with mass about $m_s$, although in
a flavor nonsinglet state.
\label{VECTOR_MASS_TABLE}
}
\end{center}\end{table}

\begin{table}[ptbh!]
\begin{center}
\begin{tabular}{|llllll|}
\hline
$am_{valence}$ & $am_{sea}$ & $am_{PV}$ & range & $\chi^2/D$ & conf. \\
\hline
0.015 ($a_1$)   & $\infty$      & 0.720(40)     & 9--25  & 11/11 & 0.48 \\
0.03 ($a_1$)    & $\infty$      & 0.730(6)      & 7--25  & 10/13 & 0.67 \\
0.015 ($b_1$)   & $\infty$      & 0.741(22)     & 6--25  & 7.3/14        & 0.92 \\
0.03 ($b_1$)    & $\infty$      & 0.748(10)     & 7--25  & 15/13 & 0.33 \\
\hline
0.01 ($a_1$)    & 0.01/0.05     & 0.820(40)     & 6--15  & 5.3/6 & 0.50 \\
0.01 ($a_1$)    & 0.01/0.05     & 0.848(24)     & 6--17  & 6.4/8 & 0.60 \\
0.007 ($a_1$)   & 0.007/0.05    & 0.767(21)     & 5--15  & 10/7  & 0.16 \\
0.005 ($a_1$)   & 0.005/0.05    & 0.790(40)     & 5--15  & 9.6/7 & 0.21 \\
0.01 ($b_1$)    & 0.01/0.05     & 1.020(90)     & 6--22  & 15/13 & 0.32 \\
0.01 ($b_1$)    & 0.01/0.05     & 0.980(60)     & 6--19  & 10/10 & 0.42 \\
0.007 ($b_1$)   & 0.007/0.05    & 0.810(40)     & 5--18  & 11/10 & 0.34 \\
0.005 ($b_1$)   & 0.005/0.05    & 0.700(90)     & 6--15  & 5.8/6 & 0.44 \\
\hline
0.031 ($a_1$)   & 0.031/0.031   & 0.667(4)      & 8--25  & 11/12 & 0.56 \\
0.0124 ($a_1$)  & 0.0124/0.031  & 0.600(8)      & 8--30  & 22/19 & 0.30 \\
0.0062 ($a_1$)  & 0.0062/0.031  & 0.532(19)     & 10--26 & 14/13 & 0.36 \\
0.031 ($b_1$)   & 0.031/0.031   & 0.681(5)      & 7--25  & 21/13 & 0.08 \\
0.0124 ($b_1$)  & 0.0124/0.031  & 0.632(9)      & 7--33  & 34/23 & 0.07 \\
0.0062 ($b_1$)  & 0.0062/0.031  & 0.650(50)     & 10--27 & 11/14 & 0.65 \\
\hline
\end{tabular}
\caption{Pseudovector meson masses.
Runs tabulated and the format are the same as in Table~\protect\ref{PS_MASS_TABLE}.
\label{PSEUDOVECTOR_MASS_TABLE}
}
\end{center}\end{table}

\subsection{Baryons}

\begin{figure}[ptbh!]
\resizebox{6.0in}{!}{\includegraphics{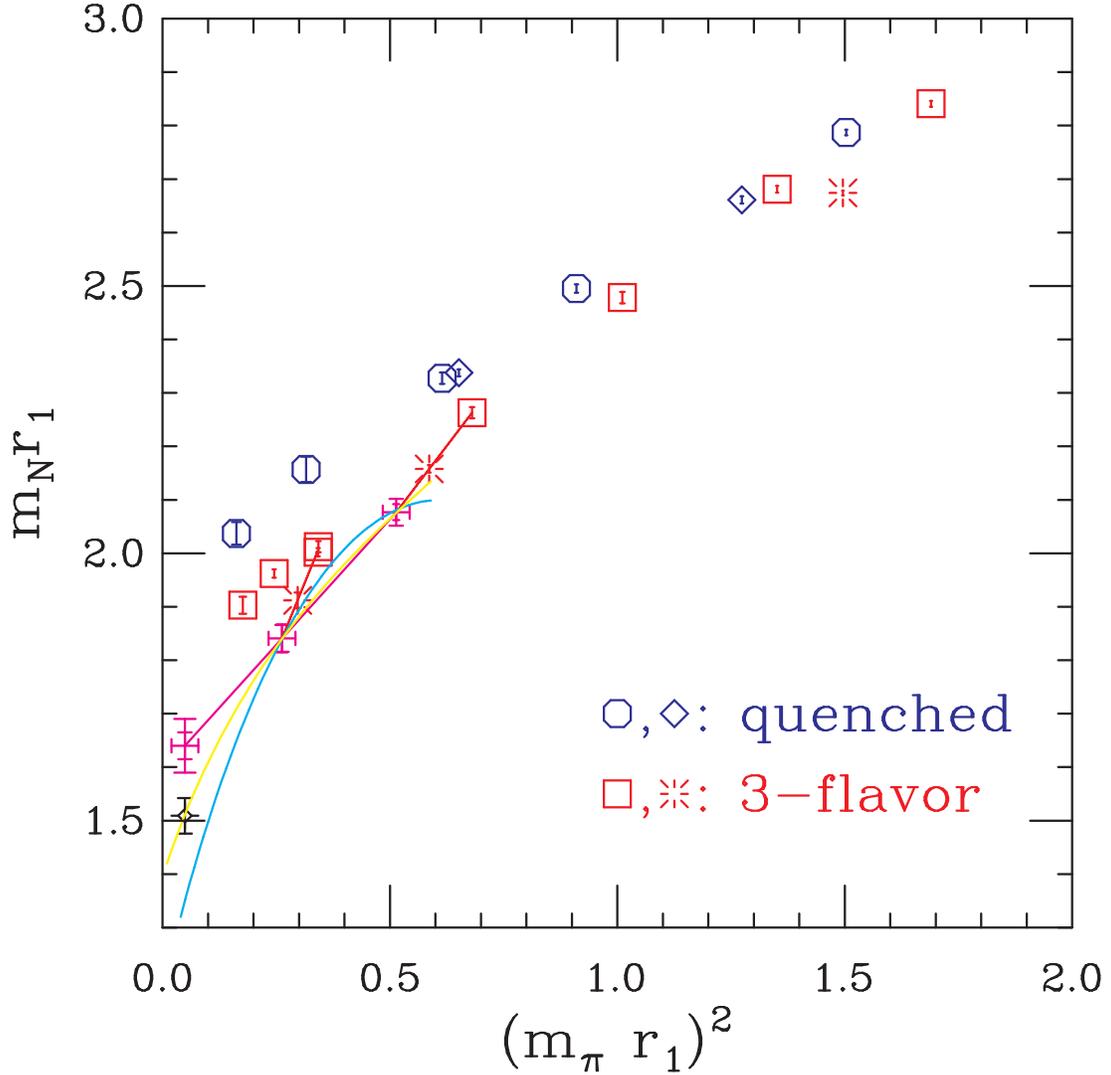}}\\
\caption{\label{MNUC_FIG}
Nucleon masses.  The \tmpblue diamonds and octagons are quenched coarse and
fine runs respectively.  The \tmpred squares are three flavor coarse lattice
results, and the \tmpred bursts the three flavor fine lattices.
The \tmpmagenta fancy plusses connected by the straight line and the
two curved lines are continuum
and chiral extrapolations discussed in the text.
The fancy diamond is the experimental value, with an error bar from the
uncertainty in $r_1$.
}
\end{figure}

Table~\ref{BARYON_MASS_TABLE} contains masses for the octet nucleon and $\Xi$.
We do not tabulate the $\Lambda$ and $\Sigma$ since our code does not cleanly
separate the light quark isospins.
In principle, the nucleon mass could be fit by methods similar to those used
for the pion mass and decay constant, incorporating effects of continuum
chiral corrections, lattice artifacts like taste symmetry breaking, finite
size effects and partial quenching.  Such an analysis is not yet available.
However, statistical errors on the nucleon mass are much larger than for the
pseudoscalars, so this full machinery may be less important here.  An alternative
strategy for dealing with lattice artifacts is to perform a continuum extrapolation
at the quark masses used in simulations, and then fit these extrapolated masses
to continuum chiral perturbation theory.  
Figure~\ref{MNUC_FIG} shows the nucleon masses in units of $r_1$.  This graph
also contains a very rough sketch of how such a continuum and chiral extrapolation
might begin.  The right most \tmpmagenta fancy plus is a linear extrapolation in $a^2 \alpha_s$
of the coarse and
fine results at $ m_{u,d} \approx 0.4 \, m_s$ to $a=0$, as indicated by the \tmpred
line.  The middle fancy plus is a similar continuum extrapolation at $ m_{u,d} \approx 0.2 \, m_s$.
The solid straight line is a linear extrapolation to the physical pion mass.
As a rough estimate of the effects of chiral logarithms, the
two curved lines are chiral perturbation theory forms constrained to match the
two continuum extrapolated points.  These forms have two free parameters, so
we emphasize that this is not a fit and there is no test of consistency of
these forms with our data.  The \tmpyellow upper curved line is an expansion
in powers of $M_\pi$ up to order $M_\pi^2 \log(M_\pi)$ from Ref.~\cite{LEINWEBER}
and the \tmpcyan lower curve is a form where the nucleon-delta mass splitting
is also treated as small\cite{VBERNARD}.
It is clear that fine lattice results at a smaller quark mass will be needed,
since the slopes of the chiral perturbation theory forms are clearly 
different from the lattice results for quark masses as small as $0.4\,m_s$.

\begin{table}[ptbh!]
\begin{center}
\begin{tabular}{|llllll|}
\hline
$am_{valence}$ & $am_{sea}$ & $am_{B}$ & range & $\chi^2/D$ & conf. \\
\hline
0.015 (N)       & $\infty$      & 0.6267(18)    & 8--23  & 14/12 & 0.30 \\
0.03 (N)        & $\infty$      & 0.7134(18)    & 12--30 & 11/15 & 0.78 \\
\hline
0.01 (N)        & 0.01/0.05     & 0.7710(40)    & 6--16  & 4.4/7 & 0.74 \\
0.01 (N)        & 0.01/0.05     & 0.7670(30)    & 6--17  & 3.8/8 & 0.88 \\
0.007 (N)       & 0.007/0.05    & 0.7480(30)    & 5--14  & 5/6 & 0.54 \\
0.005 (N)       & 0.005/0.05    & 0.7230(60)    & 5--14  & 9.8/6 & 0.13 \\
0.01/0.05 ($\Xi$)       & 0.01/0.05     & 0.9810(30)    & 8--20  & 5.3/9 & 0.81 \\
0.01/0.05 ($\Xi$)       & 0.01/0.05     & 0.9737(20)    & 8--21  & 16/10 & 0.09 \\
0.007/0.05 ($\Xi$)      & 0.007/0.05    & 0.9670(50)    & 10--20 & 9.2/7 & 0.24 \\
0.005/0.05 ($\Xi$)      & 0.005/0.05    & 0.9540(80)    & 10--19 & 4.6/6 & 0.60 \\
\hline
0.031 (N)       & 0.031/0.031   & 0.6996(11)    & 7--37  & 50/25 & 0.0023 \\
0.0124 (N)      & 0.0124/0.031  & 0.5815(19)    & 10--29 & 17/16 & 0.41 \\
0.0062 (N)      & 0.0062/0.031  & 0.5190(40)    & 11--23 & 4.6/9 & 0.87 \\
0.0124/0.031 ($\Xi$)    & 0.0124/0.031  & 0.6696(17)    & 13--33 & 12/17 & 0.80 \\
0.0062/0.031 ($\Xi$)    & 0.0062/0.031  & 0.6519(18)    & 12--30 & 19/15 & 0.21 \\
\hline
\end{tabular}
\caption{Octet baryon masses.
Runs tabulated and the format are the same as in Table~\protect\ref{PS_MASS_TABLE}.
\label{BARYON_MASS_TABLE}
}
\end{center}\end{table}

\section{Tests of Systematic and Statistical Errors}

The results in the previous two sections allow us to make several algorithm
tests as well as more physical tests.

\subsection{Single versus double precision}

As the valence quark masses are made smaller, the condition number of the fermion
matrix increases and one might worry that double precision is necessary for computing
the hadron propagators.  In general, we have used single precision for the computations
at each lattice site, with global sums in double precision.
At our smallest quark mass, $a m_{u,d}=0.005$, we have tested the accuracy of our
hadron spectrum and static potential computations by repeating the computation in double precision on
a subset of the lattices.  Table~\ref{PRECISION_TABLE} shows results for a number
of quantities evaluated on a set of 137 lattices with $a m_{l/s}=0.005/0.050$.  Note
that since these are fit on exactly the same
sets of lattices with exactly the same programs, any discrepancies
are the result of the different precision.  However, we provide
statistical errors to show how the
effects of roundoff compare with the statistical errors.
For all of these quantities the effects of using single precision are
small compared with the statistical errors, and with the statistical errors
we would get from any reasonable lengthening of this run.


\begin{table}[ptbh!]
\begin{center}
\begin{tabular}{|l|l|l|l|}
\hline
Quantity & Double & Single & Comment \\
\hline
$V(2,0,0)$ & 0.829883(852) & 0.829888(853)	&   potential at r=(2,0,0) \\
$V(2,2,2)$ & 1.05426(503)	& 1.05451(502)	& \\
$V(3,3,3)$ & 1.2511(194)	& 1.2511(194)	& \\
$r_1/a$	& 2.63933(1679)	& 2.63915(1678)	&   t=4--5, block=5 \\
$r_0/r_1$	& 1.4566(64)	& 1.4566(64)	&      t=4--5, block=5 \\
\hline
$\pi(20)$	& 411.53(1.55)	& 411.44(1.55)	&  prop. at d=20 \\
$\rho(6)$	& 143.76(1.78)	& 143.73(1.78)	& \\
$aM_\pi$	& 0.15965(22)	& 0.15966(21)	&     d=20--31, $\chi^2/D=0.60$ \\
$aM_K$	& 0.36519(34)	& 0.36519(34)	&     d=20--32, $\chi^2/D=0.82$ \\
$aM_\rho$ & 0.5330(83)	& 0.5330(83)	&      d=6--14, $\chi^2/D=0.85$ \\
$aM_N$	& 0.7311(84)	& 0.7312(84)	&      d=6--14, $\chi^2/D=0.50$\\

\hline
\end{tabular}
\end{center}
\caption{Comparison of results with single and double precision computations.
The first three lines are the static quark potential at three different
spatial separations.   These separations are in the spatial region used
in fitting the potential.
The next two lines are parameters extracted from fitting the potential,
the inverse lattice spacing in units of $r_1$ and a ``shape parameter''
$r_0/r_1$.
The second part of the table contains hadron propagator comparisons.
The $\pi(20)$ and $\rho(6)$ show the pion and rho propagators summed over
a time slice at time separations 20 and 6.  These distances
are near the minimum of the range used in fitting the masses, and so are
among the most important distances in our fits.  Finally, the last four lines are hadron
masses computed from the double and single precision propagators.
\label{PRECISION_TABLE}
}
\end{table}

\subsection{Integration step size}

Our simulation algorithm is expected to introduce errors proportional to
$\epsilon^2$ where $\epsilon$ is the simulation time step size.  Based on
previous experience and our expectations about the scaling of the fermion
force with the quark mass, we have used a step size of about 2/3 of the light
quark mass in these runs.  As a check on these effects, we have made short
runs with larger step sizes at one of our small quark masses (the same
parameters at which we checked effects of the spatial size of the lattice.)
The production runs here were done at a step size of $\epsilon=0.0067$ (658 lattices),
and the short tests at step sizes of $0.01$ (49 lattices) and $0.01333$ (53 lattices)
with lattice size $20^3\times 64$.
Table~\ref{STEPSIZE_TABLE} shows results for the static quark potential and
some hadron masses at these different step sizes, using the same fitting
ranges in each case.
Since the short runs were too short for a good error analysis, statistical errors on
these quantities are estimated by scaling the errors on the $L=20$, $\epsilon=0.0067$
run by the square root of the ratio of the numbers of configurations used.

\begin{table}[ptbh!] \begin{center} \begin{tabular}{|l|l|l|l|l|}
\hline
 & $L=20$ &  $L=28$ & $L=20$ & $L=20$ \\
Q. & $\epsilon=0.0067$ &  $\epsilon=0.0067$ & $\epsilon=0.0100$ & $\epsilon=0.013$ \\
\hline
$\Box$          & 1.70092(2)    & 1.70094(3)    &  1.70096(7) & 1.70066(7) \\
$\bar\psi\psi$  & 0.07421(10)   & 0.07420(13)   &  0.07374(37) & 0.07488(35) \\
$r_1/a$         & 2.598(8)      & 2.621(9)      &  2.649(29) & 2.619(28) \\
$aM_\pi$         & 0.22439(20)   & 0.22421(12)   &  0.22500(73) & 0.22554(70) \\
$aM_\rho$        & 0.569(5)      & 0.568(3)      &  0.557(18) & 0.558(18) \\
$aM_N$           & 0.771(4)      & 0.767(3)      &  0.785(15) & 0.753(14) \\
\hline
\end{tabular} \end{center}
\caption{
Effect of integration step size.  These are from runs with $10/g^2=6.76$ and
$a m_{l/s}=0.01/0.05$.  Columns two and three are our long runs with
$L=20$ and $28$ using a step size of $0.0067$.  (Our usual practice is to use
a step size about 2/3 of the lightest quark mass.)  Columns four and five
are from short runs with step sizes $0.01$ and $0.01333$.
\label{STEPSIZE_TABLE} } \end{table}

\subsection{Spatial size of the lattice}

In one of our coarse lattice runs, $10/g^2=6.76$, $a m_{l/s}=0.01/0.05$, we
have made a second run at a larger spatial lattice size, $28^3\times 64$.
(We have also lengthened the run with $L=20$, so this is the run where we have
the best statistics.)
This allows us to explicitly check the effects of the spatial lattice
size.  Table~\ref{SPACESIZE_TABLE} shows the results of this test for
the static quark potential and simple hadron propagators.
Note that these values of $r_1/a$ fall on opposite sides of the
interpolated (``smoothed $r_1$'') value of 2.610,
and the values of $r_0/r_1$ fall on opposite sides of a
straight line fit to the coarse lattice points
in Fig.~\ref{R0_OVER_R1_PLOT}, leading us to believe that we do not
see any statistically significant finite size effects in either the
potential or the hadron masses.
The sizes of these two lattices in physical units are 2.43 and 3.40 fm,
using $r_1=0.317$ fm to set the scale, and $M_\pi L$ is 4.48 and 6.27 respectively.  
Using the (staggered) chiral fits \cite{MILC_fpi,MILC_fpi_in-prep}
to light pseudoscalar masses and decay constants, it is possible to estimate
the leading finite volume correction on $M_\pi$.  We expect a difference
$\Delta=0.00026$ between $L=20$ and $L=28$ results, consistent with the
observed value in the simulations, $\Delta=0.00018(23)$, shown in
Table~\ref{SPACESIZE_TABLE}.

\begin{table}[ptbh!] \begin{center} \begin{tabular}{|l|l|l|l|}
\hline
Quantity 	& $L=20$ & $L=28$ & $\Delta$ \\
\hline
$r_1/a$ & 2.598(8)      & 2.621(9)      & -0.023(12) \\
$r_0/r_1$	& 1.4461(36)	& 1.4533(34)	& -0.0072(50) \\
\hline
$M_\pi$         & 0.22439(20)   & 0.22421(12)   & 0.00018(23) \\
$M_\rho$        & 0.569(5)      & 0.568(3)      & 0.001(6) \\
$M_N$           & 0.771(4)      & 0.767(3)      & 0.004(5) \\
\hline
\end{tabular} \end{center}
\caption{Comparison of results with different spatial sizes.  These
are from the runs with $10/g^2=6.76$ and $a m_{l/s}=0.01/0.05$.  The spatial
sizes were $L=20$ and $28$, corresponding to physical
sizes of $2.4$ and $3.4$ fm, using $r_1=0.317$ fm to set
the physical scale.
The first two lines are parameters extracted from fitting the potential,
the inverse lattice spacing in units of $r_1$ and a ``shape parameter''
$r_0/r_1$.
The second part of the table contains hadron mass comparisons.
$\Delta$ is the $L=20$ value minus the $L=28$ value in each row.
\label{SPACESIZE_TABLE} } \end{table}

\subsection{Autocorrelations}

\begin{figure}[ptbh!]
\resizebox{6.0in}{!}{\includegraphics{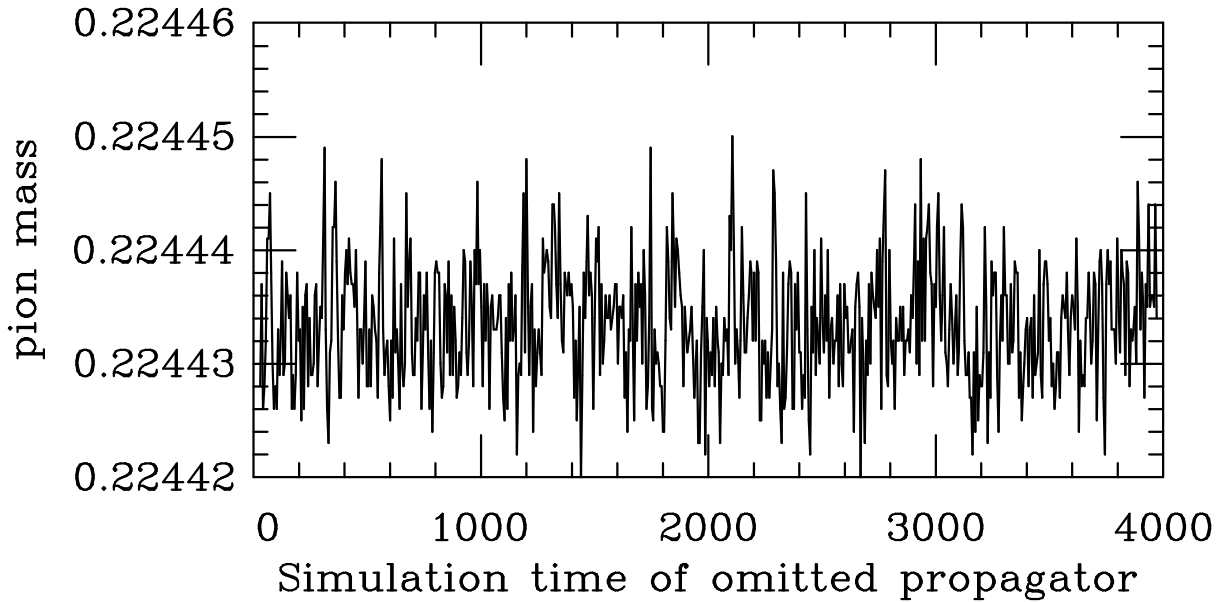}}\\
\vspace{-2.5in}
\caption{\label{pimasshistory_fig}
Single elimination jackknife masses for the pion, from the
run with $10/g^2=6.76$ and $a m_{l/s}=0.01/0.05$, using fits
with $D_{\rm min}=19$.
}
\end{figure}

\begin{figure}[ptbh!]
\resizebox{6.0in}{!}{\includegraphics{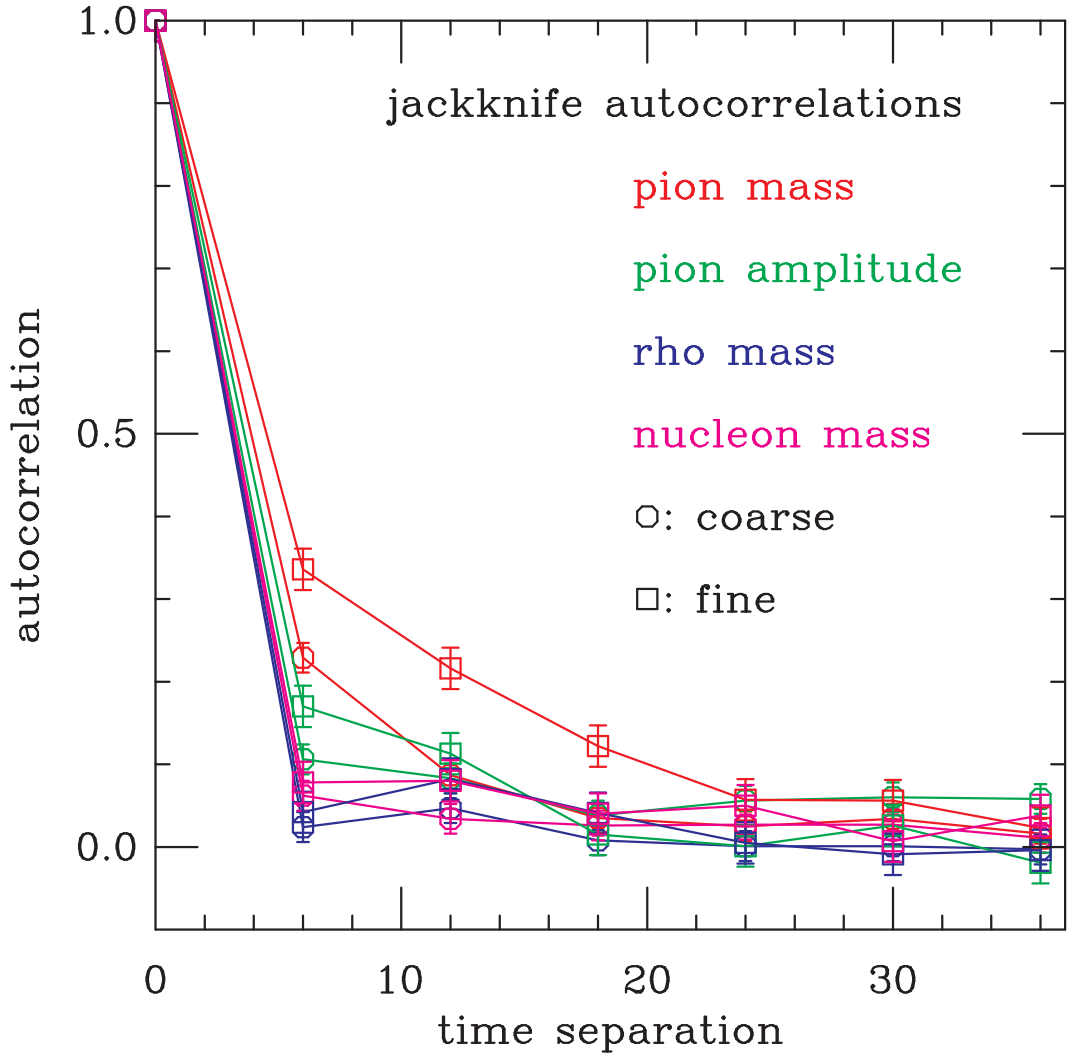}}\\
\caption{\label{autocor_mass_fig}
Normalized autocorrelations for the $\pi$, $\rho$ and nucleon
masses and the amplitude of the pion propagator as functions
of the separation in simulation time.  Results for all the coarse
lattice runs are averaged together, as are all the fine lattice runs.
}
\end{figure}

Because of the high cost of generating sample configurations with dynamical
quarks, successive samples were taken at simulation time intervals such that
they are not completely statistically independent.  The resulting autocorrelations
(in simulation time) affect the statistical errors on all of the computed
quantities.   The ``exponential autocorrelation time'', which
is determined by the eigenvalue of the Markov process matrix which is closest to one,
is expected to be the same for all calculated quantities.
However, the contribution of this slowest mode to various quantities varies,
and to parameterize the effect of autocorrelations on individual quantities
we use the ``integrated autocorrelation time'', $\tau_{\rm int} = \sum_{s} C_Q(s)$, where
$s$ runs over the simulation time separations and $C_Q(s)$ is the normalized
autocorrelation for quantity $Q$,
\BE C_Q(s) = \frac{\langle Q(t+s) Q(t) \rangle - \langle Q \rangle^2}
{\langle Q(t) Q(t) \rangle - \langle Q\rangle^2}\ \
\ .\EE
Because we need a covariance matrix to calculate masses from the average
propagators, and getting a nonsingular covariance matrix requires more
samples than there are points in the fit range, we cannot get a hadron
mass from one sample.  So, to study autocorrelations of hadron mass estimates
we use the ``mirror image'' of this procedure --- we do single elimination
jackknife fits with one sample omitted from the data set and compute the
autocorrelations of these jackknife fits.
Figure~\ref{pimasshistory_fig} shows the jackknife pion masses
as a function of the simulation time of the omitted sample for the run
with $10/g^2=6.76$ and $a m_{l/s}=0.01/0.05$.
For example, Table~\ref{autocor_6_table} shows $C_Q(6)$ where $Q$ is
the $\pi$, $\rho$ or nucleon mass or the amplitude in the pion propagator,
and the simulation time separation is six units, corresponding to successive
stored lattices.  From this table we can see that the normalized autocorrelation
is largest for the pion mass, and has no obvious systematic dependence on the
light quark mass.  Therefore, we average the autocorrelations over the quark
masses, separately for the coarse and fine runs.   The resulting autocorrelations
as a function of simulation time separation are plotted in Fig.~\ref{autocor_mass_fig}.


Not surprisingly, the autocorrelation times are larger on the fine lattices
than on the coarse lattices.  In Ref.~\cite{MILC_topology} autocorrelations
of the topological charge were computed on these lattices.  The topological charge
evolves more slowly than the hadron masses, with estimated autocorrelation
times as large as 35 time units for the $10/g^2=7.18$, $a m_{l/s}=0.031/0.031$ run.
We refer the reader to \cite{MILC_topology} for more discussion.

\begin{table}[ptbh!] \begin{center} \begin{tabular}{|c|l|l|l|l|l|l|}
\hline
$10/g^2$ & $a m_{u,d}$ & N & $M_\pi$ & $A_\pi$ & $ M_\rho$ & $M_N$ \\
6.85 & 0.05 & 425 & 0.196 &  0.079 &  0.047 &  0.077 \\
6.83 & 0.04/0.05 & 351 & 0.383 &  0.127 &  -0.031 & 0.119 \\
6.91 & 0.03/0.05 & 564 & 0.274 &  0.161 &  0.082 &  0.070 \\
6.79 & 0.02/0.05 & 486 & 0.173 &  0.169 &  0.025 &  0.143 \\
6.76 & 0.01/0.05 & 658 & 0.229 &  0.056 &  0.046 &  0.014 \\
6.76 & 0.007/0.05 & 487 & 0.150 &  0.056 &  -0.055 & -0.020 \\
  & average & 2971 & 0.229 &  0.106 &  0.024 &  0.062 \\
\hline
7.18 & 0.031 & 496 & 0.426 &  0.223 &  0.074 &  0.203 \\
7.11 & 0.0124/0.031 & 534 & 0.311 &  0.142 &  -0.002 & 0.034 \\
7.09 & 0.0062/0.031 & 586 & 0.283 &  0.152 &  0.055 &  0.011 \\
 & average &  1616 & 0.336 &  0.170 &  0.042 &  0.078 \\
\hline
\end{tabular} \end{center}
\caption{
Normalized autocorrelations $C_Q(6)$ for hadron masses and the
pion amplitude in the light quark runs.  The third column is the
number of samples in each run.   We also show the results averaged
over all the coarse runs, and over all the fine runs, where the third
column is the total number of coarse or fine lattices.
\label{autocor_6_table} } \end{table}

\section{Hadronic decays and excited states}

\begin{figure}[ptbh!]
\resizebox{6.0in}{!}{\includegraphics{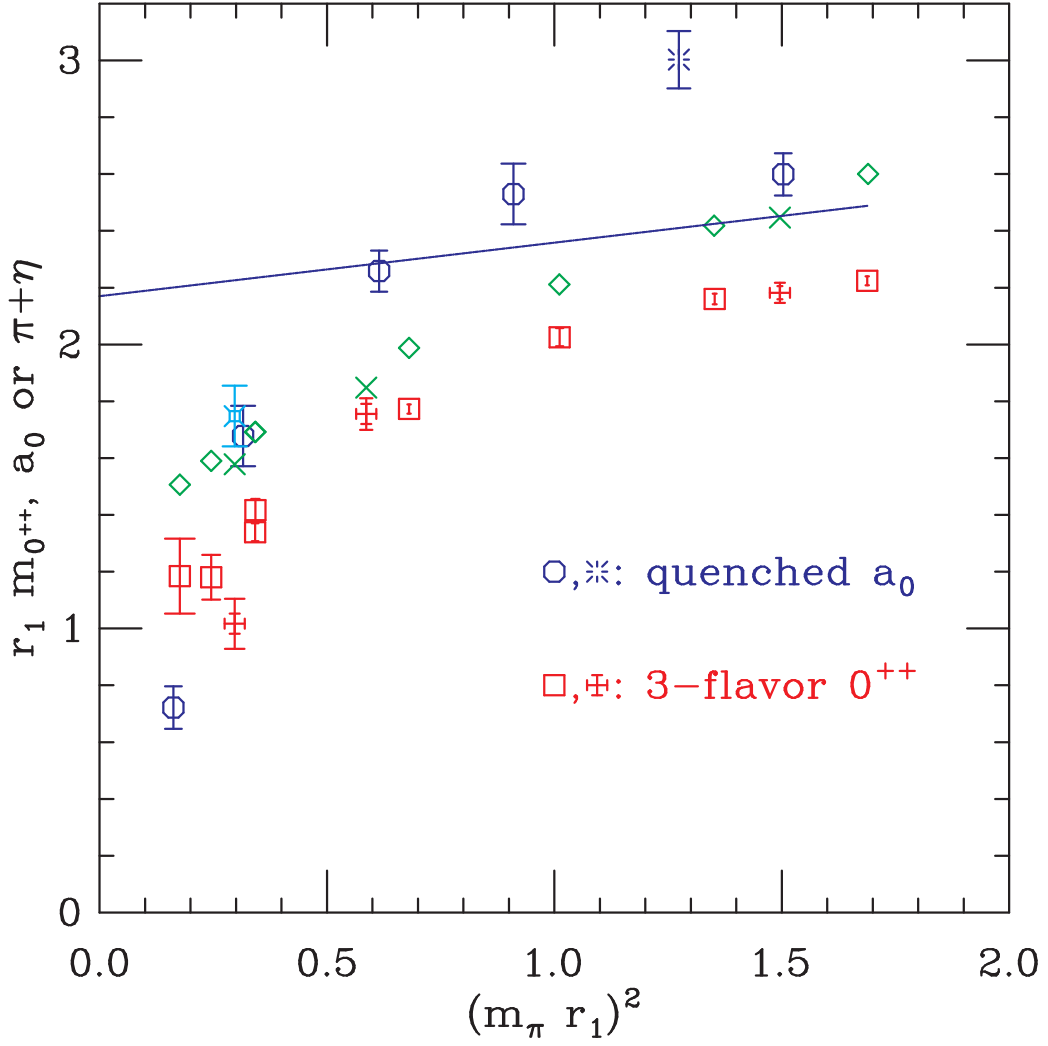}}
\caption{\label{A0_DECAY_FIG}
$0^{++}$ energies.
The \tmpred squares are three flavor coarse runs, and the \tmpred fancy plusses,
three flavor fine runs.  The \tmpblue octagons are a quenched coarse run and the
\tmpblue burst a quenched fine run.
The \tmpcyan decorated square is an excited $0^{++}$ mass from one of the runs.
The \tmpgreen diamonds and crosses are
sums of $\pi$ and $\eta$ masses on coarse and fine runs respectively, where
the $\eta$ mass is estimated from $M_\eta^2=\frac{1}{3} M_\pi^2 + \frac{2}{3} m_{s\bar
s}^2$, with $m_{s\bar s}$ the unmixed $s \bar s$ pseudoscalar mass.
The straight \tmpblue line is an extrapolation of $a_0$ masses
from heavier quark runs (not shown in this graph).
}
\end{figure}

When the quark mass is small enough, most of the hadrons we study
are unstable, decaying strongly into two or more lighter hadrons.
In principle, although not always in practice, fitting to the ground
state mass in our propagators will give the mass of the lightest
state with the right quantum numbers in the periodic box, which in
many cases will be a two particle state.
In Ref.~\cite{MILC_spectrum1} we showed this effect in the $0^{++}$
($a_0$) channel.  Figure~\ref{A0_DECAY_FIG} updates this plot with
more results on coarse lattices at light quark mass, and the new
results on the fine lattices.  For the three flavor runs, the fine
lattice points agree well with the coarse lattice results.
The figure also shows the mass of the lowest energy two-meson
state expected to couple to this particle, $\pi+\eta$.
Surprisingly, the new points at the lighest quark masses
increasingly deviate from this two-meson mass, which is not
understood.
The light mass quenched propagators remain difficult to fit, which
may not be surprising for unstable particles in an
unphysical theory.
We have not yet tried fitting to the particle-plus-ghost form
suggested by Bardeen {\it et al.}\ \cite{Bardeen:2001jm}.
For quark masses where the two-meson state has lower energy, it would
be satisfying to find a one meson ($a_0$) state as an excited state in
the propagator.  Our attempts to do this have been unsuccessful so far.
In the fine lattice run at $a m_{l/s}=0.0062/0.031$ we were able
to extract an excited state mass, shown as the \tmpcyan decorated square
in Fig.~\ref{A0_DECAY_FIG}.   However, the mass of this state is
still much smaller than the extrapolations from large quark
mass, and it is likely also a two-meson state, perhaps $K \bar K$.

\begin{figure}[ptbh!]
\resizebox{6.0in}{!}{\includegraphics{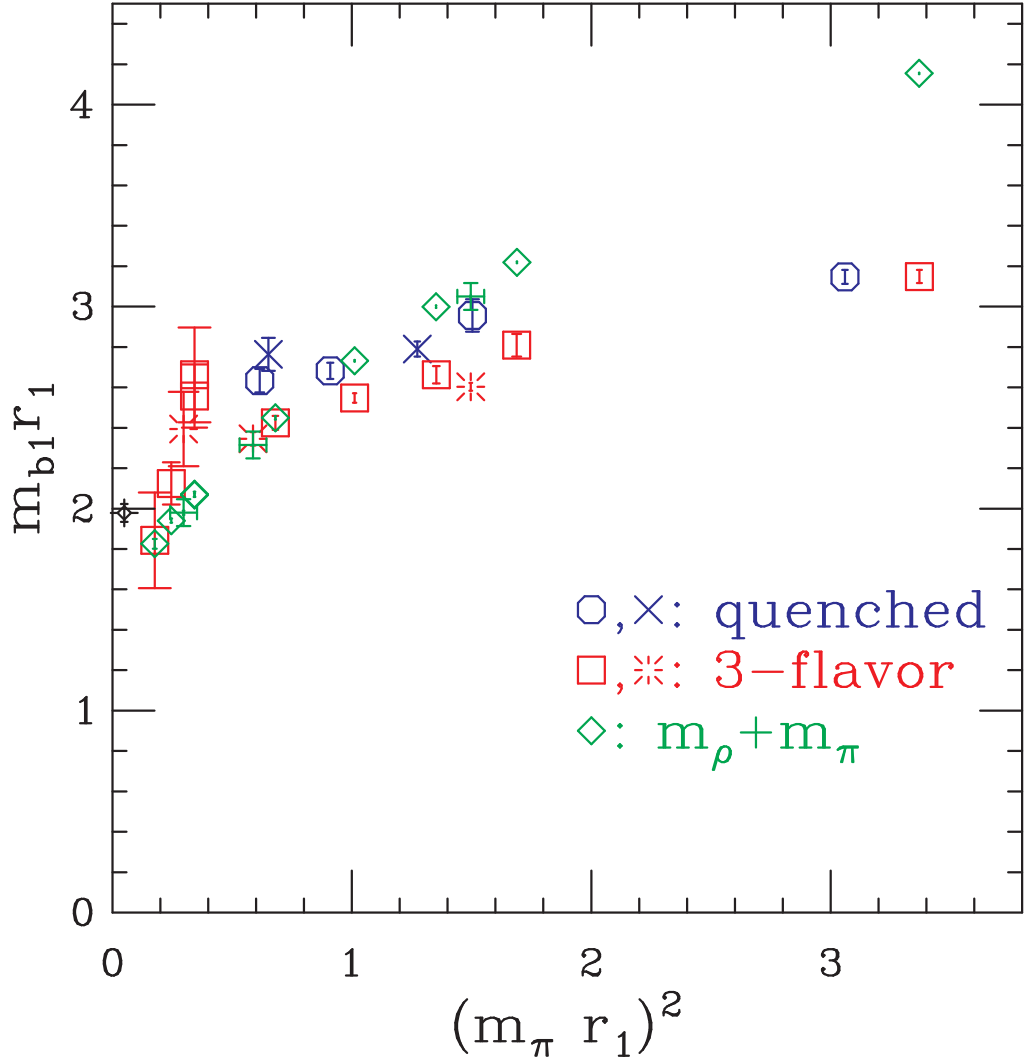}}
\caption{\label{MB1_R1_FIG}
$1^{+-}$ energies.
The \tmpred squares are three flavor coarse runs and the \tmpred bursts
three flavor fine runs.  The \tmpblue octagons and crosses are quenched
coarse and fine runs respectively.  The \tmpgreen diamonds and fancy plusses are
sums of rho and pion masses on coarse and fine three flavor runs respectively.
The fancy diamond on the left is the experimental value, with an error
bar corresponding to the uncertainty in $r_1$.
}
\end{figure}

We also expect to see the pseudovector ($1^+$) mesons couple to
two zero-momentum mesons, although for these mesons we are not
as far below the threshold as in the $0^{++}$ case.
Figure~\ref{MB1_R1_FIG} shows $1^{+-}$ ($b_1$)
masses as a function of quark mass along with the decay channel mass
$M_\rho+M_\pi$.  We tentatively attribute the downturn at the lightest
quark masses to this decay, although better statistics
at the lightest coarse lattice and a lighter mass fine lattice run
would clarify the situation.  Again, we are unable to get good fits
for the lightest mass quenched propagators.

\begin{figure}[ptbh!]
\resizebox{6.0in}{!}{\includegraphics{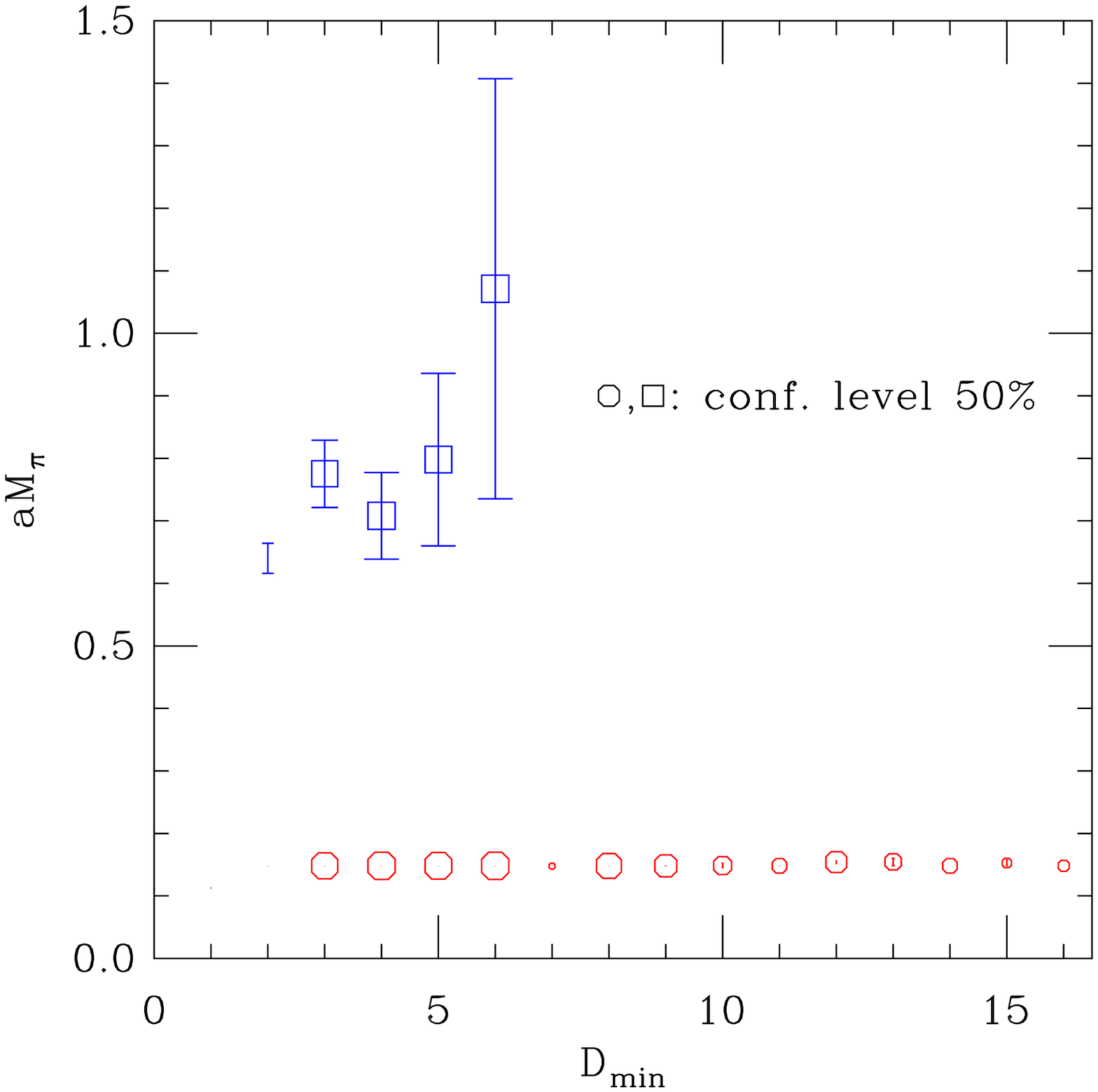}}
\caption{\label{pion_fit_plot}
Fit plot showing ground state and excited state masses of Goldstone pion 
$0^{-+}$ as a function of $D_{\rm min}$, the minimum distance included in the 
fit.   This is from the run with $10/g^2=7.09$, $am_{l/s}=0.0062/0.031$, with fits
using $D_{\rm max}=28$. The
symbol size is proportional to confidence level.
}

\end{figure}

\begin{figure}[ptbh]
\resizebox{6.0in}{!}{\includegraphics{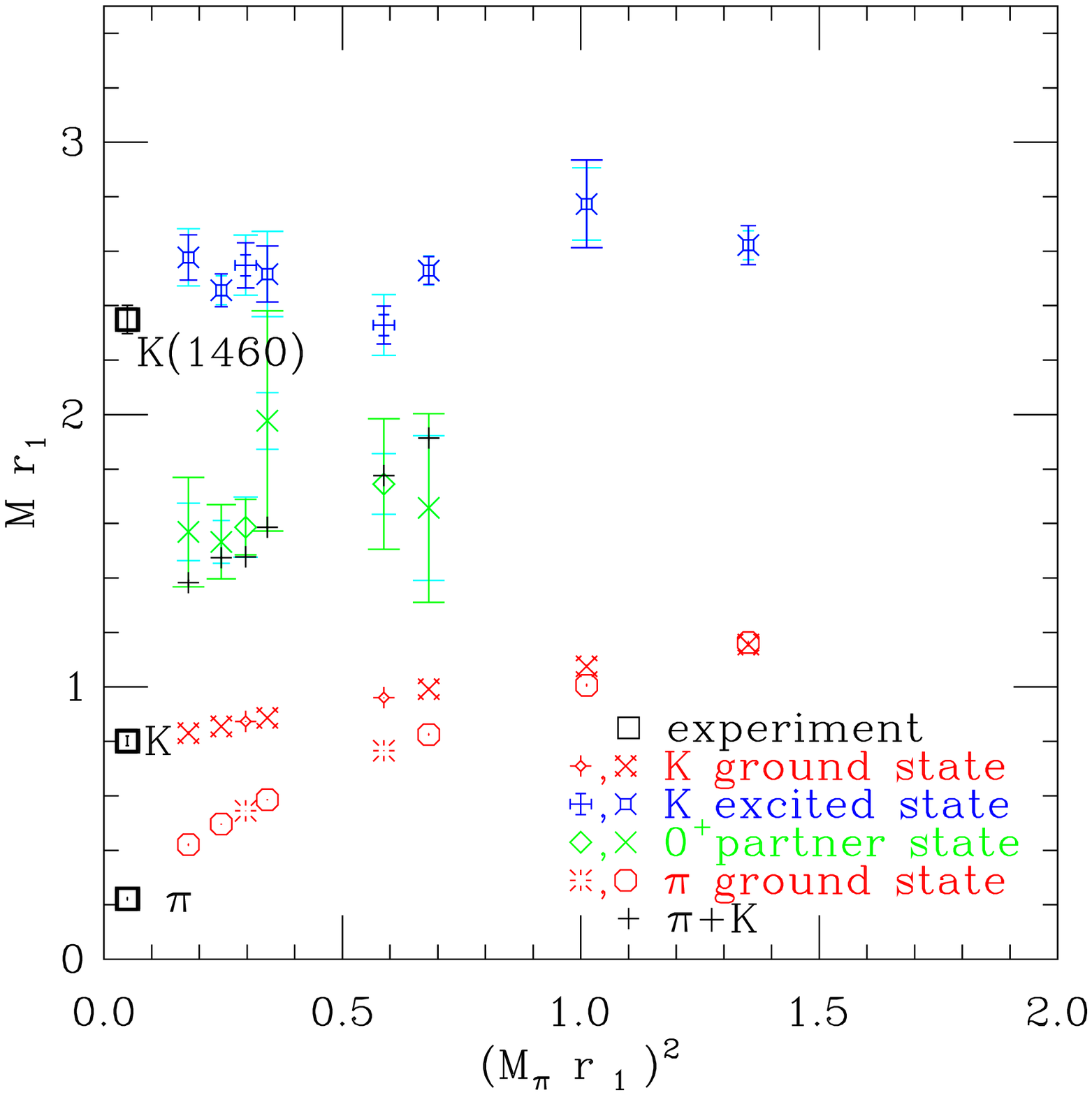}}
\caption{\label{kaon_summary}Summary of fits of kaon propagators. Ground
state and excited state kaon masses are interpolated to the correct strange
quark mass. $0^+$ parity partner state and $\pi+K$ masses are {\em uncorrected}
for comparison. For the $K$s, $\pi$ and $0^+$ entries in the legend, the symbol
on the left represents fine lattice results, and that on the right the coarse 
lattice results.
}
\end{figure}

Kogut-Susskind meson propagators generally include normal exponential 
contributions from one $J^{PC}$ value and an oscillating exponential
component from a parity partner state. In the case of the Goldstone pion, 
the parity partner has the exotic $J^{PC}=0^{+-}$ and thus does not 
contribute to the propagator. In combination with a relatively high 
signal-to-noise ratio at all time separations, this enhances our ability 
to determine the $0^{-+}$ contributions. Specifically, in addition to the 
one-state fits, which we presented in Figure \ref{dmin_ps_0062_fig} and 
Table \ref{PS_MASS_TABLE}, when we performed a two-state fit of the 
pseudoscalar propagator data, we were able to determine the mass of a 
second, excited $0^{-+}$ state. We have presented preliminary results of 
this analysis in 
\cite{MILC_excited}.
We fit $0^{-+}$ propagators to the form:
\begin{equation}
\label{2normal_states}
C(t)=A_0(e^{-M_0t} + e^{-M_0(T-t)}) 
+A_1(e^{-M_1t} + e^{-M_1(T-t)}),
\end{equation}
where $A_0$ and $M_0$ are the amplitude and mass of the ground state, and
$A_1$ and $M_1$ are the same for the lowest excited state. Figure 
\ref{pion_fit_plot} is a sample pion fit plot showing the fitted values of 
$aM_0$ and $aM_1$ as a function of the minimum time separation, $D_{\rm min}$, 
included in the fits. By comparing to one-state fits shown in 
Figure \ref{dmin_ps_0062_fig}, note the inclusion of an excited state in 
the fitting function allows high-confidence fits to extend down to a 
$D_{\rm min}$ of 2 or 3, as might be expected. The excited state's contribution
to the propagator decays to unresolvable levels relatively quickly, however, 
and consequently larger fit distances are often not so useful.
\begin{figure}[ptbh!]
\resizebox{6.0in}{!}{\includegraphics{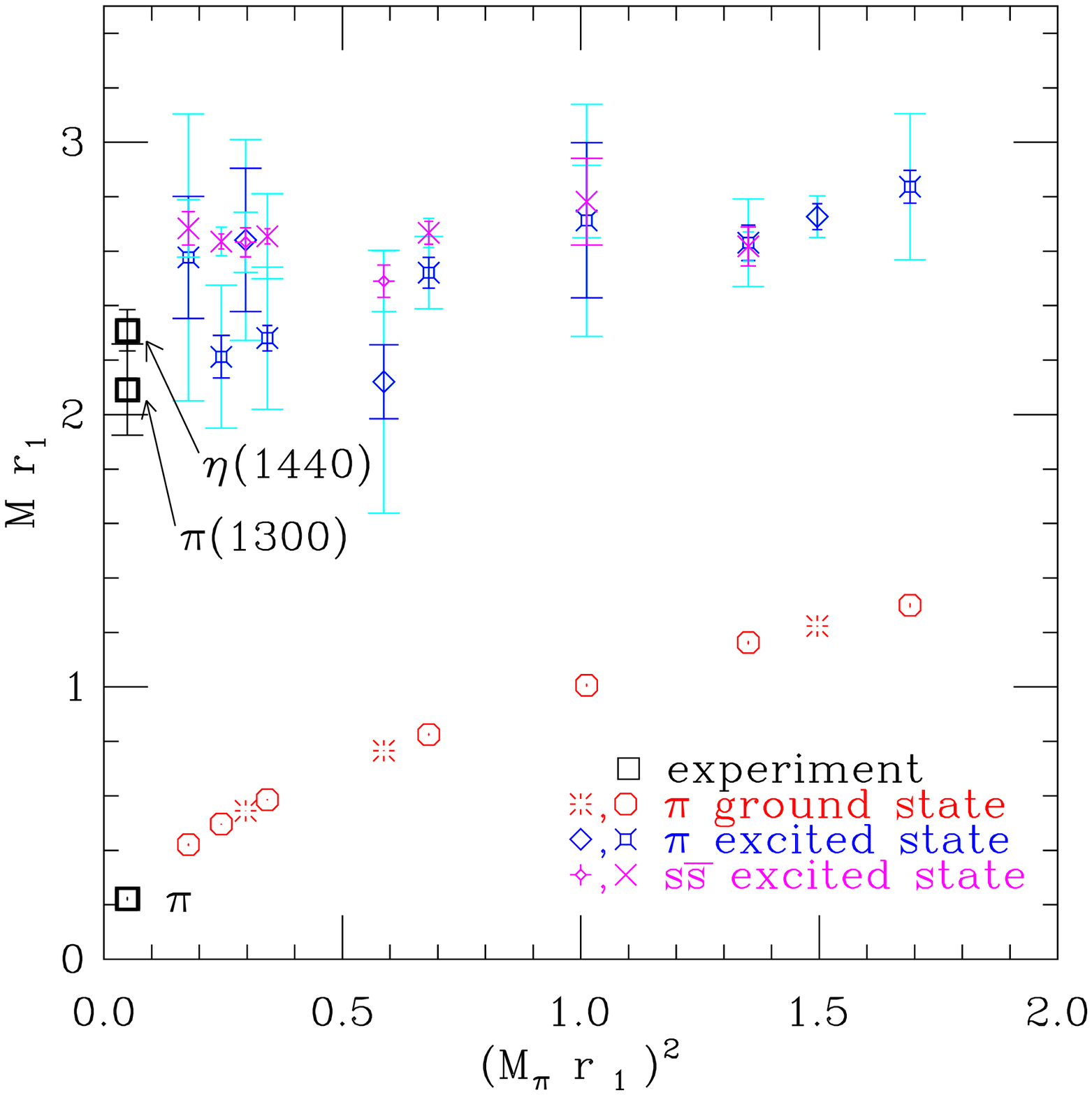}}
\caption{\label{pion_fits}
Ground state and excited state $0^{-+}$ masses as a function of 
$(M_\pi r_1)^2$. In the legend, the symbol on the left represents fine 
lattice results, and that on the right the coarse lattice results.
}
\end{figure}
Figure \ref{pion_fits} summarizes the two-state fits for the $0^{-+}$ 
masses as a function of $(M_\pi r_1)^2$. These excited state masses fit
a linear function of $(M_\pi r_1)^2$ to a 12\% confidence level.
As the statistical errors on the excited pion mass fits
are large
compared with the differences between the coarse and fine lattice fits, we 
considered all of the mass fits together in the linear fit.
Extrapolating the resulting  linear function to the physical value of
$(M_\pi r_1)^2=0.050$, we get a prediction of a physical $0^{-+}$ excited 
state at  1362(41)(247)MeV, which agrees within the large errors with the
mass of the $\pi(1300)$ state.
The first error is statistical. The second is the systematic error 
predominantly due to contributions to the propagator which are unaccounted 
for in the form of the fitting function. We estimate this by examining the
fit plots and estimating the range of mass values one might reasonably 
choose, that is, this error reflects the stability of the fitted value under
variation of the fit range, e.g., the difference between the $D_{min}=3$ and 
$D_{min}=4$ points in Figure \ref{pion_fit_plot}
and is reflected in Figures \ref{pion_fits} and \ref{kaon_summary} as 
light \tmpcyan error bars on the excited states. We linearly extrapolate the 
individual systematic errors to $(M_\pi r_1)^2= 0.05$.  Systematic errors due
to chiral extrapolation, finite lattice size and lattice spacing, are 
small relative to the statistical error and the systematic
error from additional states.

Similarly, an excited state is evident in the $0^{-+}$ $s\overline{s}$  
propagator.
The analysis of states containing strange quarks is complicated by the fact 
that our simulated strange quark masses, $am_s= 0.050, 0.031$ differ from the 
physical strange quark mass, $am_s^*=0.039, 0.027$ (for the coarse and fine 
lattices respectively) as discussed in subsection \ref{ps_mesons}. To correct 
for this,  after fitting to  the form of Eq. (\ref{2normal_states}),
we interpolated the meson masses to the correct physical values of the strange 
quark mass, $m_s^*$, using
\begin{equation}
\label{m_s_interpolation}
M_{\rm PS}(m_s^*) 
= M_{\rm PS}(m_s) - (m_s - m_s^*)\frac{M_{\rm PS}(m_s) - M_{\rm PS}(m_{u,d})}{m_s - m_{u,d}},
\end{equation}
where we use the mass of the excited $0^{-+}$ state at the simulation value of 
$m_s$ 
for $M_{\rm PS}(m_s)$, and the pion excited state  on the same lattices
for $M_{\rm PS}(m_{u,d})$. We cannot interpolate masses from lattices with 
three flavors of degenerate quarks in this manner, so we eliminate them from 
this analysis. 

The interpolated excited state masses fit a linear function
of $(M_\pi r_1)^2$ and we again extrapolated the resulting form to the physical
$(M_\pi r_1)^2$.  The result is $M_{s\overline{s}}= 1645(40)(145)$ for the 
excited $s\overline{s}$ psuedoscalar state.

We have no pure $s\overline{s}$ physical $0^{-+}$ 
with which to compare ground state fits. We can, however, compare the 
extrapolation of the corrected excited state masses 
with the experimental mass of the $\eta(1440)$, which one expects to be 
dominated by the $s\overline{s}$ contribution. 
This is consistent with our result with the large systematic error.
We display all of the pion and (corrected) $0^{-+}$ $s\overline{s}$ fits in 
Figure \ref{pion_fits}, with physical states for comparison.

\begin{figure}[ptbh!]
\resizebox{6.0in}{!}{\includegraphics{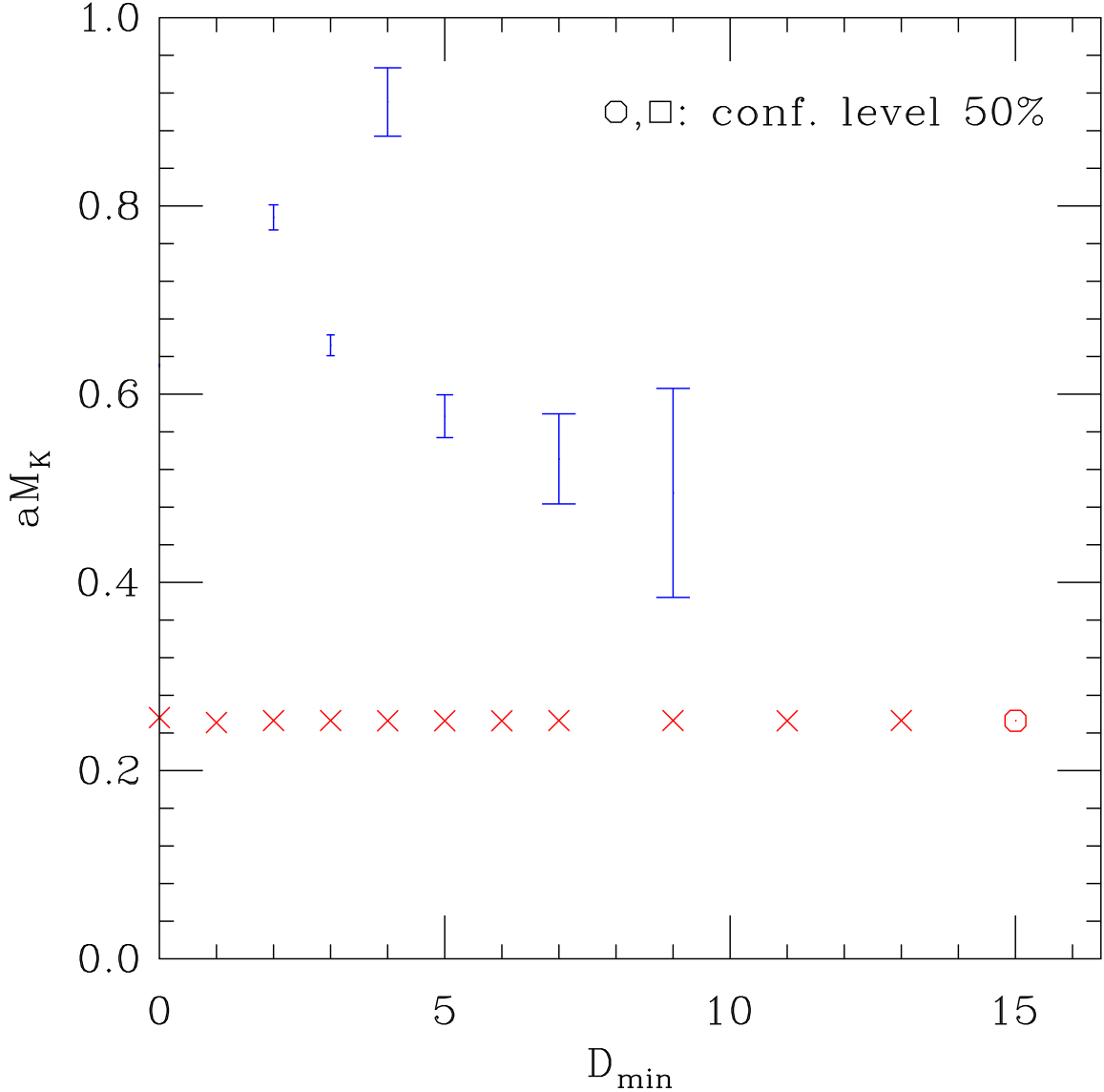}}
\caption{\label{kaon_2state_fits}
Two-state fits to three-flavor kaon pseudoscalar propagators as a 
function of minimum 
distance included in the fit
from the run with $10/g^2=7.09$ and $a m_{l/s}=0.0062/0.031$.
The size of the symbols is proportional to the confidence level of the 
fit. Octagons and squares represent the two $0^-$ states, although, as
discussed in the text, all of the confidence levels for this fit
are so low that these symbols are extremely small.
Standard size crosses are used for points where both the error bar and the confidence
level are too small to be visible otherwise.
These fits used $D_{\rm max}=30$.
}
\end{figure}

\begin{figure}[ptbh!]
\resizebox{6.0in}{!}{\includegraphics{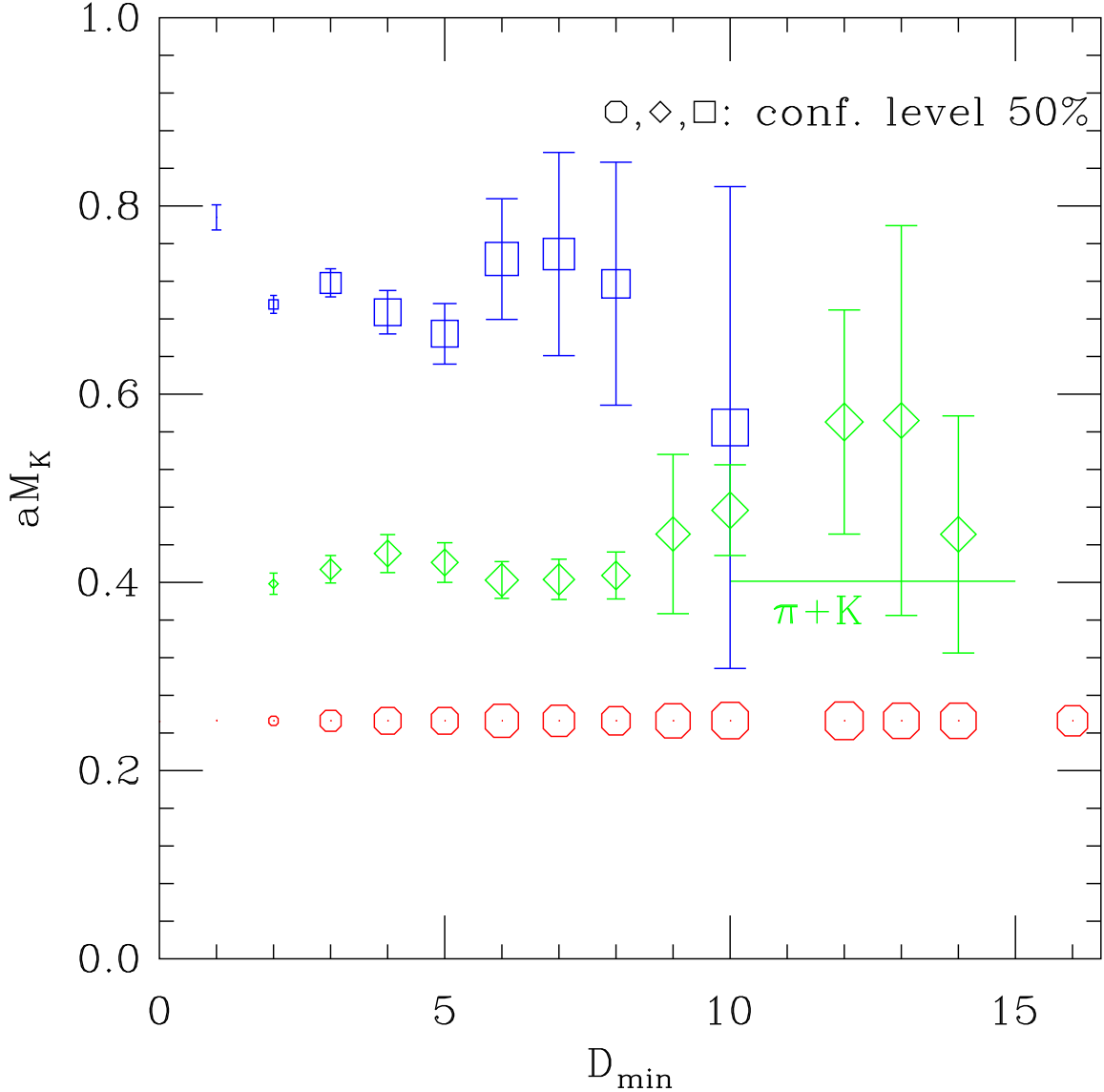}}
\caption{\label{kaon_3state_fits}
Three-state fits to three-flavor kaon pseudoscalar propagators as a 
function of minimum 
distance included in the fit
from the run with $10/g^2=7.09$ and $a m_{l/s}=0.0062/0.031$.
The size of the symbols is proportional to the confidence level of the 
fit. Octagons and squares represent the two $0^-$ states; diamonds represent the 
oscillating $0^+$ state. $D_{\rm max}=30$.
}
\end{figure}

Even more interesting is the kaon propagator. Formed of a light quark and 
a strange quark, the kaon, $J^{P}=0^-$, has no definite charge-conjugation 
quantum number when $ m_{u,d}\neq m_s$. Consequently, it has a non-exotic 
parity partner with $J^{P}=0^+$, and the propagator has a tiny, but 
significant oscillating component. On theses lattices the amplitude of the 
oscillating state 
is significantly smaller than that of the kaon ground state, and the 
mass is greater than that of the kaon ground state, thus it does not 
interfere with with single-exponential fits of the propagator at large 
time separations ($D_{\min}>14$).  Two-state fits to the form of 
Eq.~(\ref{2normal_states}) fail at all time separations because the $0^+$ mass 
falls below that of the first excited $0^-$ state. 
Figure \ref{kaon_2state_fits} shows an attempt to fit the $10/g^2=7.09$,
$am_{u,d}=0.0062$, $am_s=0.031$ fine lattice propagator to two non-oscillating 
exponentials, as in Eq.~(\ref{2normal_states}). All fits are 
of extremely low confidence levels and there is no evident plateau for
the excited state.
Figure \ref{kaon_3state_fits} shows fits of the same propagator to a 
three-state form,
\begin{equation}
\label{3_states}
C(t)=A_0(e^{-M_0t} + e^{-M_0(T-t)}) 
+A_1(e^{-M_1t} + e^{-M_1(T-t)})+A_2(-1)^t(e^{-M_2t} + e^{-M_2(T-t)}),
\end{equation}
with high confidence levels and masses of consistent value through a large variation 
in the lower limit of the fit range, $D_{\rm min}$.

Propagators from both fine lattice sets with $m_{u,d}\neq m_s$ were 
inconsistent with double exponential forms, (Eq.~\ref{2normal_states}),
but fit to triple exponentials, (Eq.~\ref{3_states})
with high confidence. The same was true of the  coarse lattice sets with 
$am_{u,d}\leq 0.02$. In general, we find that
as $ m_{u,d} \longrightarrow m_s$, the amplitude of 
the oscillating state becomes indistinguishable from zero, presumably because
charge conjugation regains its status as a good quantum number. In the fits
to  kaon propagators from  the coarse lattice set with
$10/g^2=6.79$, $am_{u,d}=0.030$ 
we were no longer able to distinguish the amplitude of any oscillating state 
from noise. Confidence levels for both two-state and three-state fits were a 
few tenths of a percent, yet we could discern equivalent plateaus for the 
excited $0^-$ state mass as a function of $D_{\rm min}$ in each case. 
Attempts to read a plateau for the oscillating state were unconvincing. For 
the coarse lattices with $10/g^2=6.81$, $am_{u,d}=0.040$, and both coarse and 
fine lattices with three degenerate flavors of quarks, two state fits resolved
the excited state with high confidence (as we have mentioned before when we 
considered these very same fits as limiting cases of both pions and 
$0^{-+}$ $s\overline{s}$ mesons.)

\begin{figure}[ptbh]
\resizebox{6.0in}{!}{\includegraphics{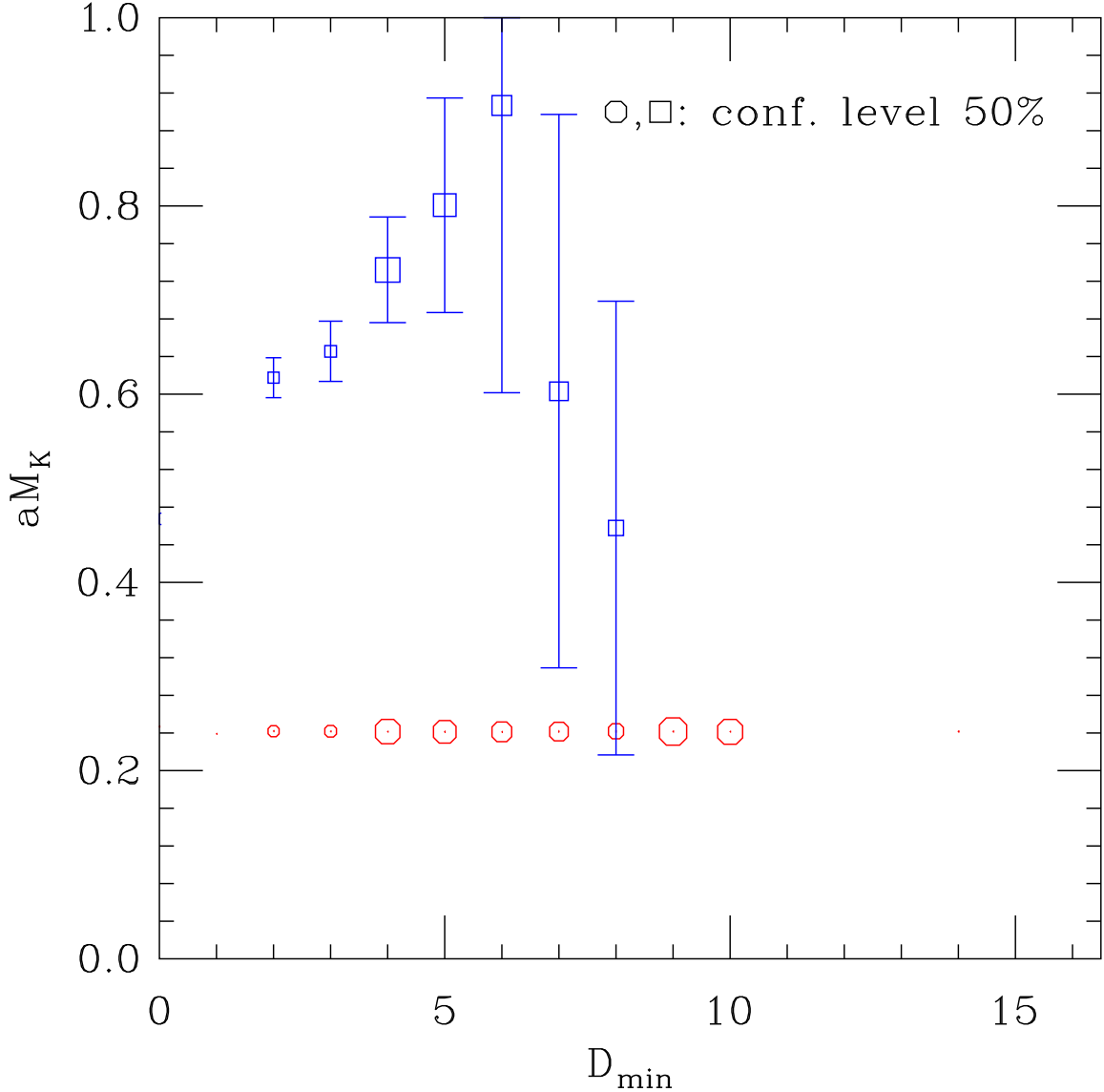}}
\caption{\label{quenched_kaon_ex}Quenched kaon mass fit plot showing 
ground state and first excited state, with $10/g=8.40$, 
$a m_{l/s}=0.0062,0.031$,
and $D_{\rm max}=17$. Fit to Eq.~(\ref{2normal_states}) {\em without} 
an oscillating state.
}

\end{figure}

The oscillating scalar state is far lighter than the lightest strange 
$0^+$ meson, the  $K_0^*(1430)$. It does, however, agree well with the sum 
of the masses of the dominant $K_0^*(1430)$ decay mode products, $K+\pi$, on
every lattice set for which it was measured.
Resolution of the $K_0^*(1430)\rightarrow K+\pi$ decay channel
is additional evidence that our simulations with light dynamical quarks
correctly reproduce the expected complexities of the physical world.
When we perform similar fits to quenched kaon propagators we can find no
evidence of an oscillating $0^{+}$ state, even with widely separated valence 
quark masses, such as $a m_{l/s}=0.0062/0.031$. Furthermore, with the quenched
kaon propagators, it is simple to extract the contribution of the first 
excited
$0^{-}$ state, see for example Fig. \ref{quenched_kaon_ex}.

We have also performed an extrapolation of the excited kaon state masses to 
the physical value of $(M_\pi r_1)^2=0.050$. Again considering the
fine and coarse lattice data together the excited states fit,
with 8\% confidence level, to a line which intercepts $(M_\pi r_1)^2=0.050$
at 1527(46)(68)MeV. This is in decent agreement with the K(1460) state and 
inconsistent with the K(1460)'s expected decay products, $\pi\pi K$, which
should be at about 775 MeV. This lends credence to the belief that the K(1460)
is a true mesonic state.

Figure \ref{kaon_summary} summarizes the fits to the kaon propagators. As
with the $s\overline{s}$ states, we have corrected the ground state and 
excited state mass fits for the difference between the simulated strange 
quark mass and the physical strange quark mass using the interpolation 
expression (\ref{m_s_interpolation}). Since we have measured a $0^+$ state
at only one value of the strange quark mass for each lattice spacing, 
interpolation of the $0^+$ state is not possible. We include the pion
ground state and the sum of the pion and (uncorrected) kaon ground state 
masses for comparison. We include isospin-averaged physical states for 
comparison. We display these results numerically in Table \ref{zero_minus_fits}.

\begin{table}[ptbh!] \begin{center} \begin{tabular}{|c|c|c|c|c|c|c|c|c|}
\hline
$10/g^2$ & $a m_{u,d}/a m_{s}$ & $N_{\rm states}$ &  $a M_{0^+}$ &  $A_{0^+}$ & $aM_\pi+aM_K$ & $aM_{K_{\rm ex}}$ & range & conf\\
\hline
6.85 & 0.05      & 2 & --- & ---           & 0.97 & 1.05(2)(10) & 3-18 & 0.36\\
6.83 & 0.04/0.05 & 2 & --- & ---            & 0.90 & 1.02(3)(2) & 4-32 & 0.36\\
6.81 & 0.03/0.05 & 2 & ---           & ---  & 0.81 & 1.07(3)(5) & 4-26 &0.008\\
6.79 & 0.02/0.05 & 3 & 0.63(12)(10) & $-3(2)$ & 0.72 & 0.96(3)(2) & 3-16 & 0.39\\
6.76 & 0.01/0.05 & 3 & 0.76(15)(4) & $-13(9)$ & 0.61 & 1.00(5)(6) & 3-16 & 0.27\\
6.76 & 0.007/0.05 & 3 &0.58(4)(3)  & $-10(2)$ & 0.56 & 0.97(3)(3) & 3-16 & 0.28\\
6.76 & 0.005/0.05 & 3 &0.60(6)(4)  & $-24(6)$ & 0.53 & 0.98(3)(4) & 3-21 & 0.56\\ 
\hline
7.18 & 0.031 & 2 & --- & ---               & 0.64 & 0.71(1)(4) & 5-25 & 0.83 \\
7.11 & 0.0124/0.031 & 3 & 0.47(6) &  $-7(3)$ & 0.48 & 0.64(2)(3) & 5-30 & 0.49\\
7.09 & 0.0062/0.031 & 3 & 0.43(2) & $-22(3)$ & 0.40 & 0.69(2)(3) & 4-30 & 0.64\\
\hline
\end{tabular} \end{center}
\caption{Results of two- and three-state fits to $0^-$ kaon propagators.
\label{zero_minus_fits} } \end{table}


It is worth pointing out that we fit these excited state masses in wall 
source propagators that were designed specifically to minimize the 
contribution of excited states.  It is likely that analysis with other 
quark sources would further enhance our ability to resolve excited states.

We note that the consistency of the excited $K$ and $\pi$ states
with experiment indicates that there is no unphysical scale in these channels
of length $\gtwid 2$ lattice spacings.  This is encouraging, since
non-localities that might be introduced by taking the fourth root of the
staggered determinant could show up here.

\section{Conclusions}

\begin{figure}[ptbh!]
\resizebox{6.0in}{!}{\includegraphics{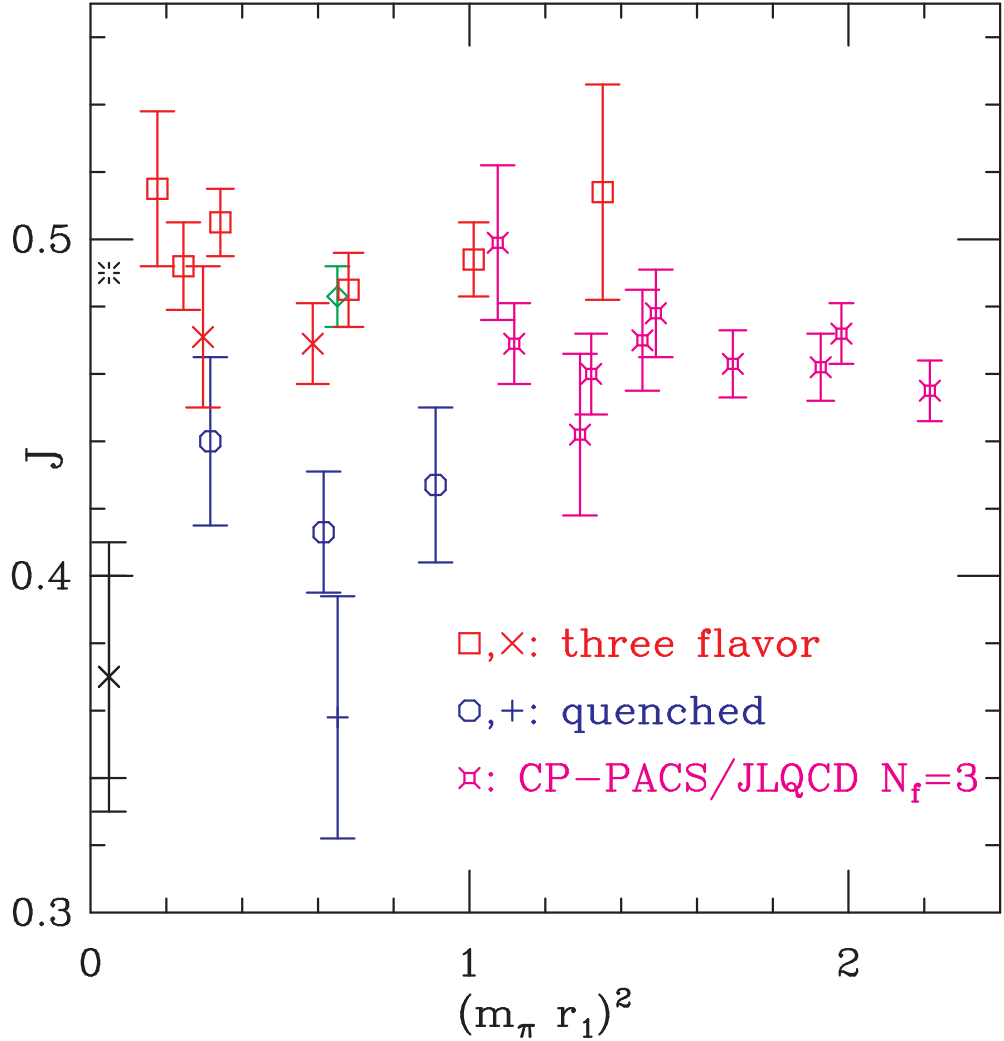}}
\caption{\label{J_fig}
The ``J'' parameter.  The \tmpred squares and crosses are three flavor coarse
and fine lattice results respectively.  The \tmpblue octagons and plus signs
are quenched coarse and fine results, while the \tmpgreen diamond is a two
flavor run.  The \tmpmagenta decorated squares are CP-PACS/JLQCD three flavor
Wilson quark results\protect\cite{Kaneko2003}.   The cross at the left is
the original UKQCD quenched estimate\protect\cite{UKQCD_J}, and the burst at
the left is the experimental value.
}
\end{figure}

In this project we have calculated hadron masses including the effects
of three flavors of dynamical quarks, using light quark masses down
to $0.1 \, m_s$ and lattice spacings of about $0.12$ and $0.09$ fm.
These quark masses are light enough that we are beginning to ``see
hadronic decays'' in the sense that the lowest energy states for some
quantum numbers may be two-meson states instead of a single particle.
To the extent that we can reasonably expect, our spectrum results are
consistent with the experimental hadron spectrum.  One quantity that is
sensitive to the effects of sea quarks is ``$J$'', which is roughly the
derivative of the vector meson mass with respect to the squared pseudoscalar
mass\cite{UKQCD_J}.   In particular, we plot
\BE J = \frac{M_{K^*} \LP M_\phi - M_\rho \RP}{2\LP M_K^2-M_\pi^2\RP} \ \ \ .\EE
This quantity is plotted in Fig.~\ref{J_fig}, which updates results
from \cite{MILC_spectrum1}, and also includes recent points from the
CP-PACS/JLQCD collaboration\cite{Kaneko2003}.

Comparison of lattice results with the physical spectrum still requires extrapolations
to zero lattice spacing and to the physical quark masses.
In principle, the extrapolation to zero lattice spacing is straightforward ---
we expect errors proportional to $a^2 g^2$.  Extrapolation to the physical
light quark mass is more difficult.  First, most of the hadrons decay
strongly, and as we have seen for the $0^{++}$, and the $0^+$ for
nondegenerate quarks, simulations with light sea quark masses show
the couplings to the decay channels.   For stable hadrons the
extrapolation to physical light quark mass involves chiral logarithms.
Because of the remaining breaking of taste symmetry, fitting to the
chiral logarithms requires that the continuum extrapolation be done
first, or simultaneously.

In the case of the pseudoscalar masses and decay constants, taste violations
have been included in the chiral perturbation theory, which makes possible
a simultaneous extrapolation in lattice spacing and quark masses\cite{MILC_fpi, MILC_fpi_in-prep}.
The small statistical errors on pseudoscalar masses and decay constants
make this rather involved analysis necessary, but also make it possible.
Work towards comparable extrapolations for some other quantities, such
as the nucleon mass, is in progress.

\begin{figure}[ptbh!]
\resizebox{6.0in}{!}{\includegraphics{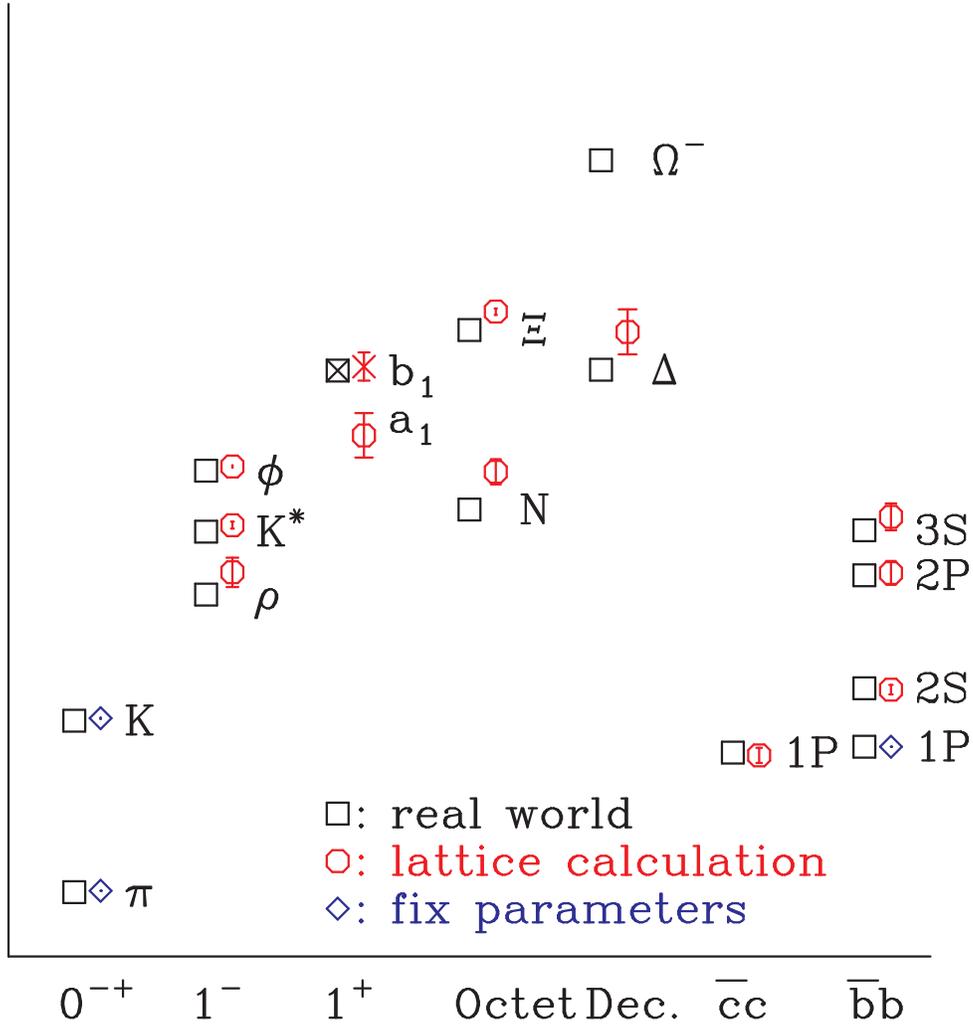}}
\caption{\label{bigpic_fig}
The ``big picture''.  Crude continuum and chiral extrapolations
of hadron masses and splittings compared with experimental values.
The upsilon and charmonium columns are differences from the
ground state masses, from work of the HPQCD and Fermilab groups\cite{HPQCD_private,PRL}.
Here the $\pi$ and $K$ masses fix the light and strange quark masses,
and the $\Upsilon$ 1P-1S mass splitting is used to fix the lattice spacing.
}
\end{figure}

In the meantime, it is interesting to use a less sophisticated extrapolation
to see how
these lattice results compare with the real world.
Figure~\ref{bigpic_fig} shows such a comparison, using a
linear or quadratic extrapolation in the light quark mass and linear 
extrapolation in the squared lattice spacing.   Since the difference
between the strange quark mass used in our simulations and the
correct value is roughly twice as large in the coarse runs as in the
fine runs, the extrapolation in lattice spacing also largely corrects
for the too-large strange quark mass used in the runs.
(It is not entirely an accident that the continuum extrapolation largely
takes care of adjusting the strange quark mass, since one of the largest reasons
for the error in adjusting the strange quark mass was the neglect of
order $a^2$ corrections in tuning the strange quark mass.)
Note that the lattice nucleon mass plotted here is the linear extrapolation
shown in Fig.~\protect\ref{MNUC_FIG}; a proper chiral extrapolation is
expected to lower this value.

The spectrum results from these simulations with
three dynamical light flavors are encouraging.  Clearly, however, considerably more work is needed,
in particular on chiral extrapolations, before we can be confident that the calculations can produce
accurate and precise results in all the channels that we have examined.
Runs are continuing for $m_{u,d}=0.1 m_s$ on both coarse and fine lattices.

\section*{ACKNOWLEDGEMENTS}
Computations for this work were performed at the San Diego Supercomputer
Center (SDSC), the Pittsburgh Supercomputer Center (PSC), Oak Ridge National Laboratory (ORNL),
the National Center for Supercomputing Applications (NCSA),
the National Energy Resources Supercomputer Center (NERSC),
the Albuquerque High Performance Computing Center,
Indiana University, and the University of Arizona.
We thank Takashi Kaneko for sharing results.
This work was supported by the U.S. Department of Energy under contracts
DOE--DE-FG02-91ER-40628,      
DOE--DE-FG02-91ER-40661,      
DOE--DE-FG02-97ER-41022       
and
DOE--DE-FG03-95ER-40906       
and National Science Foundation grants
NSF--PHY99-70701              
and
NSF--PHY00--98395.            


\end{document}